\documentclass[12pt,preprint]{aastex}

\usepackage{color,graphicx}

\def\laur#1{Los Alamos Report LA-UR-#1}

\begin{document}

\setcounter{equation}{0}
\setcounter{figure}{0}

\title{A Fast Potential and Self-Gravity Solver for Non-Axisymmetric
  Disks} 

\author{Shengtai Li\altaffilmark{1},
  Matthew J. Buoni\altaffilmark{2}, and Hui Li\altaffilmark{3}}
\altaffiltext{1}{
Theoretical Division, MS B284, Los Alamos National
Laboratory, NM 87545; sli@lanl.gov}
\altaffiltext{2}{
Department of Mechanical Engineering, University of California, Santa
Barbara, CA 93107;
buoni@engineering.ucsb.edu}
\altaffiltext{3}{
Theoretical Division, MS B227, Los Alamos National
Laboratory, NM 87545; hli@lanl.gov}


\begin{abstract}

  Disk self-gravity could play an important role in the dynamic
  evolution of interaction between disks and embedded protoplanets.
  We have developed a fast and accurate solver to calculate the disk
  potential and disk self-gravity forces for disk systems on a uniform
  polar grid. Our method follows closely the method given by Chan et
  al. (2006), in which an FFT in the azimuthal direction is performed
  and a direct integral approach in the frequency domain in the radial
  direction is implemented on a uniform polar grid. This method can be
  very effective for disks with vertical structures that depend only
  on the disk radius, achieving the same computational efficiency as for
  zero-thickness disks.

  We describe how to parallelize the solver efficiently on distributed
  parallel computers. We propose a mode-cutoff procedure to reduce the
  parallel communication cost and achieve nearly linear scalability
  for a large number of processors. For comparison, we have also
  developed a particle-based fast tree-code to calculate the
  self-gravity of the disk system with vertical structure. The
  numerical results show that our direct integral method is at least
  two order of magnitudes faster than our optimized tree-code approach.

\end{abstract}


\section{Introduction}

Well before extrasolar planets were discovered, \cite{GoTr79,GoTr80}
and \cite{LiPa86a,LiPa86b} have speculated that 
tidal interactions between disks and embedded protoplanets would lead
to planet migration. \cite{Ward97} suggested that two different types of
migration could occur. \cite{NeBe03a,NeBe03b} studied numerically the 
effects of disk self-gravity in two-dimensional simulations of planet-disk
interactions. \cite{PiHu05a} reported by an analytical
derivation that the disk gravity accelerates the planetary
migration. Recently, simulations by \cite{BaMa08} confirmed that the 
self-gravity indeed accelerates the type I migration. They implemented a 
2D self-gravity solver in their code using the fast Fourier transform
(FFT) method of \cite{BiTr87},  
which requires a logarithmic radial spacing ($\log(r)$). 

In Newtonian gravity, we
can define the gravitational potential $\Psi$ associated with the mass
density, $\rho$, by the volume integral
\begin{equation}
\Psi({\bf x}) = -G\int\int\int\frac{\rho({\bf x'})}{|{\bf x}-{\bf
    x'}|}d^3x'
\label{eq_pot1}
\end{equation}
over all space, where $G$ is the gravitational constant.
Rewriting Eq. (\ref{eq_pot1}) in differential form, we obtain
Poisson's equation
\begin{equation}
\nabla^2\Psi = 4\pi G\rho,
\label{eq_pot2}
\end{equation}
with $\Psi$ satisfying the boundary condition $\Psi(\infty)= 0$ at
all times. The numerical ``Poisson solvers'' to Eq. (\ref{eq_pot2}) can be
classified into two categories: difference methods and integral
methods. The difference methods solve Eq.
(\ref{eq_pot2}) directly by either finite-difference or finite-element
methods with necessary boundary conditions. The known boundary conditions at
infinity are usually not very useful if a finite domain is considered. User
specified boundary conditions or Dirichlet boundary 
conditions obtained by direct
summation are often required.  A key advantage of difference 
methods is that, generally, they are relatively fast once
initialized. However, they have low or limited accuracy, and they
often rely on the integral method to provide the boundary conditions.

The integral methods are to integrate Eq. (\ref{eq_pot1}) directly.
They have the advantage that the summation stops naturally at the
domain boundaries. However an integral method has two difficulties in
a practical implementation. One is that the integral has a point mass
singularities (i.e. when ${\bf x'}\to {\bf x}$ in
Eq. (\ref{eq_pot1})), which is often 
circumvented by introducing softening. The other difficulty is that
they are computationally prohibitive for a large system. The
computational cost can be significantly 
lowered if the FFT method can be used.

The gravitational field, $g=-\nabla \Psi$, is more convenient in many
applications. It can be calculated via numerical
derivatives, which generally have poor precision. To obtain high
accuracy, we can integrate the field directly by
\begin{equation}
g({\bf x}) = G\int\int\int\frac{\rho({\bf x'})({\bf x'}-{\bf
    x})}{|{\bf x}- {\bf x'}|^3}d^3x'~~,
\label{eq_f1}
\end{equation}
which shares the same two difficulties as calculating
Eq. (\ref{eq_pot1}).

In this paper, we present a method for computing the disk self-gravity
for quasi-2D disk models. We consider a disk with the cylindrical
grid $(r,\phi,z)$. We assume that the scale height of the disk
(semi-thickness) is only radius-dependent, i.e., $H= H(r)$, and it is 
quite small, $H(r)/r \ll 1$ (namely geometrically thin disks). 
We also assume that the vertical structure of the
density can be described by some function $Z(r,z)$ that is independent
of $\phi$ and time $t$, i.e.,
$$
\rho(t,r,\phi,z) = \Sigma(t,r,\phi)Z(r,z).
$$
In this paper, we are particularly interested in the gravitational
potential and field force at
the $z=0$ plane. Although our method is valid for any function of
$Z(r,z)$, we assume that the density vertically has a Gaussian
distribution. Under this assumption,
\begin{equation}
Z(r,z) = \frac{1}{\sqrt{2\pi
    H(r)^2}}\exp\left(-\frac{z^2}{2H(r)^2}\right).
\label{eq_z}
\end{equation}
With the presence of the vertical structure (\ref{eq_z}), the
migration rate of the protoplanet is reduced up to 50\% (see \cite{Koller04}).
We assume that the disk has a constant sound speed $c_s$. Then from the scale
height
$$
\frac{H}{r} = \frac{c_s}{v_K}~~,
$$
where $v_K$ is the Keplerian velocity $v_K(r) = \sqrt{GM_\star/r}$, we obtain
$$
H(r) = \frac{c_s r^{3/2}}{\sqrt{GM_\star}}~~.
$$
Other profiles for the scale height are also easy to handle. For
example, if the aspect ratio, $h=H/r$, is constant, then $H(r) =
rh$. The numerical verification in \S \ref{sec:test} actually uses a constant
$H$ in the whole domain. 

Since we are interested in the potential and field in the $z=0$ plane, the
easiest method to obtain $\Psi$ is to integrate directly the equation
\begin{equation}
\Psi(t,r,\phi) = \int_{r_{\min}}^{r_{\max}}\int_0^{2\pi}\Sigma(t,r',\phi')r'dr'd\phi'
\int_{-\infty}^{\infty}-\frac{GZ(r',z')}{\sqrt{r^2+{r'}^2-
    2rr'cos(\phi-\phi')+{z'}^2}} dz'
\label{2d_eq1}
\end{equation}
where $r_{\min}$ and $r_{\max}$ are the inner and outer radial
boundaries, and $\Sigma$ is the vertically integrated density. As in 
\cite{Hure05}, we have employed two different 
coordinate systems in the radial direction: $r$ as the field grid
and $r'$ as the source grid. For simplicity, we set $G=1$ and $M_\star = 1$.

The rest of the paper is organized as follows. In \S \ref{sec:non-axi}, 
we describe the method given by \cite{ChPO06} on how to solve 
Eq. (\ref{2d_eq1}) using a direct integral via a Green's function method.
Our method, though essentially follows the one given by \cite{ChPO06},
differs in that we use a direct summation method on a uniform grid in
the radial direction, which is the 
same grid used in our hydro code. In addition, we present two approaches to
circumvent the singularity in Eq. (\ref{2d_eq1}) and two
methods to calculate the force field. For comparison, we have also implemented
a 2D tree-code with a simplified 3D treatment to calculate the self-gravity 
for disks with vertical structures. In \S \ref{sec:test} we present an
efficient parallel implementation scheme on distributed memory computers. 
We describe a new algorithm to calculate the gravity force at an
arbitrary point. We also present numerical test results to compare
different approaches. A few concluding remarks are given in 
\S \ref{sec:conc}.

\section{Green's Function Method \label{sec:non-axi}}

We present here a numerical method to compute integral
(\ref{2d_eq1}). For zero-thickness non-axisymmetric disk, the classical 
FFT method \citep[e.g.][]{BiTr87} based on polar grid, which has
been implemented by \cite{BaMa08}, is often used. However, as
pointed out by \cite{HuPi05}, the FFT method has a few
drawbacks. First it requires a grid with a logarithmic radial spacing,
which could be inconvenient to most hydro-solvers using uniform
spacing. Secondly, to avoid the well-known alias issue, the FFT method
requires to double the number of cells along the radial direction. The
FFT calculation is thus 
done on a grid that has twice the extent of the hydrodynamic grid along the
radial direction, which induces some complications in the calculation of the
convolution kernels, as well as many communications between both grids.
Furthermore, with the presence of vertical
structure, it is impossible to apply the FFT
method in 3D cylindrical coordinates because neither the potential nor
the gravitational field can be represented as convolution products in
$z$. There is no such coordinate transformation as described in
\cite{BiTr87} to make the integral (\ref{2d_eq1}) become a convolution
product in all coordinates. 

The Green's function method given by \cite{ChPO06} avoids the FFT in
the radial direction. Instead, a pseudo-spectral method on a scaled cosine
radial grid is used to achieve the high-order accuracy. Moreover, a known,
time-independent vertical structure is easy to incorporate. In the
following, we describe modifications to their method so that it can be
applied directly to a uniform radial grid.  

\subsection{Modified method given by \cite{ChPO06}}

\cite{ChPO06} proposed a direct integral method to solve Eq. 
(\ref{2d_eq1}). For the sake of completeness, we recap the key steps
in their method here. First, we introduce a softening $\varepsilon$ to
Eq. (\ref{2d_eq1}) and denote
\begin{equation}
\mathcal{G}(r,r',\phi-\phi') =
\int_{-\infty}^{\infty}-\frac{Z(r',z')}{\sqrt{r^2+{r'}^2 -
    2rr'cos(\phi-\phi')+\varepsilon^2+{z'}^2}}dz',
\label{intGz}
\end{equation}
where $\varepsilon$ can be zero or a radius-dependent parameter that
will be discussed later in Section \ref{sec:ts}.
Function $\mathcal{G}(r,r',\phi-\phi')$ can be computed
either analytically or by numerical quadrature. 
For the special case of $Z(r,z)$ defined by (\ref{eq_z}),
$$
\mathcal{G}(r,r',\phi-\phi') = -
\frac{e^{R^2/4}K_0(R^2/4)}{\sqrt{2\pi}H(r')}
$$
where $R^2 = (r^2+{r'}^2-2rr'\cos(\phi-\phi')+\varepsilon^2)/H^2(r')$,
and $K_0$ denotes the modified
Bessel function of the second kind. Let
\begin{equation}
I(r,r',\phi-\phi') = 2\pi r'\mathcal{G}(r,r',\phi-\phi').
\label{eq_I}
\end{equation}
Then Eq. (\ref{2d_eq1}) becomes
\begin{equation}
\Psi(t,r,\phi) =
\int_{r_{\min}}^{r_{\max}}\frac1{2\pi}\int_0^{2\pi}\Sigma(t,r',\phi')
I(r,r',\phi-\phi')  
dr'd\phi'
\label{2d_eq1_2}
\end{equation}
Note that both $\Sigma$ and $I$ are periodic functions with period
$2\pi$, and they are in a natural convolution representation in
Eq. (\ref{2d_eq1_2}). Applying Fourier transform to (\ref{2d_eq1_2})
with respect to (w.r.t.) $\phi$, and using the convolution theorem, we obtain
\begin{equation}
\hat\Psi_m(t,r) =
\int_{r_{\min}}^{r_{\max}}\hat\Sigma_m(t,r')\hat{I}_m(r,r')dr', \qquad
m\in[-\infty,+\infty],
\label{2d_eq3}
\end{equation}
where $\hat{f}_m$ represents the coefficients in the
Fourier series expansion of $f$, which is given as 
$$
\hat{f}_m = \frac1{2\pi}\int_0^{2\pi}f(\phi)e^{-im\phi}d\phi~~.
$$
Note that our equation (\ref{2d_eq3}) is slightly different from
the equation (39) in \cite{ChPO06}, since we have put all the
relevant coefficients in the representation of $I$ in Eq. (\ref{eq_I}). 

The integral over $r$ in Eq. (\ref{2d_eq3}) was evaluated using Chebyshev 
spectral method on a ``Chebyshev-roots grid'' by \cite{ChPO06}. 
However, for most hydro codes that use a uniform grid, it is desirable to use
the same uniform grid for both hydro and self-gravity solutions to avoid
interpolation between the hydro solver and self-gravity solver (which
is inconvenient and introduces interpolation error). 

We propose to integrate (\ref{2d_eq3}) directly with numerical
quadrature using the available discrete values of $\hat\Sigma_m$ and
$\hat{I}_m$ on a uniform grid.
Note that $I(r,r',\phi-\phi')$ has an analytic expression and is
not changed with time as long as the vertical structure remains the same. 
As proposed by \cite{ChPO06}, we can pre-compute its Fourier
transform, $\hat{I}_m(r,r')$, 
to speed up the algorithm. The whole algorithm can be summarized as
follows:
\begin{enumerate}
\item For each pair $(r,r')$, we pre-compute $\hat{I}_m(r,r')$. 
Take the Fourier transform of $I(r,r',\phi-\phi')$, which is defined in
Eq. (\ref{eq_I}), w.r.t. $\phi-\phi'$,
$$
\hat{I}_m(r,r') =
\mbox{FFT}(I(r,r',\phi-\phi')), \qquad m=1,2,...,N_\phi,
$$
where $N_\phi$ is the number of cells in $\phi$-direction.
Since $I(r,r,\phi-\phi')$ is a real and even function w.r.t. $\phi-\phi'$,
only the discrete cosine transform is needed and $N_\phi/2$ modes need to be
stored.

\item Take the Fourier transform of $\Sigma(r',\phi')$ w.r.t. $\phi'$
  to obtain $\hat\Sigma_m(r')$, $m=1,2,...,N_\phi$.
\item Calculate Eq. (\ref{2d_eq3}) by numerical quadrature in the radial
  direction to obtain $\hat\Psi_m(r)$. Either the midpoint or
  trapezoidal rule can be
  used, depending on the grid point distribution of the source grid.
\item Take the inverse Fourier transform of $\hat\Psi(r)$
  w.r.t. $\phi$ to obtain  $\Psi(r,\phi)$.
\end{enumerate}
We should emphasize that the pre-computing of $\hat{I}_m(r,r')$ plays
a major role in the efficiency of our self-gravity solver. This
step can be calculated once for all during a long time simulation of the
disk-planet interaction system. We find that it costs at least fifty
times more than the rest of steps. One cannot afford to calculate it in
every time step. 

The overall computational cost is $O(N_rN_\phi\log N_\phi+ N_\phi
N_r^2)$, where $N_r$ is 
the number of cells in the radial direction. Again, these steps are
essentially the same as in \cite{ChPO06}, except that the integration
is done on a uniform polar grid using direct summation.

\subsection{Treatment of singularity\label{sec:ts}}

When $\varepsilon=0$, the integrand in Eq. (\ref{intGz}) contains a
singularity. Note that the density splitting method of \cite{PiHu05b}
cannot be used in our case because the density
profile is unknown during the pre-computing of $\hat{I}_m(r,r')$.

We have tested two approaches to solve the singularity issue. One
approach is to use a nonzero softening, $\varepsilon(r')$. We propose a
function that scales roughly linearly with $\Delta r$, i.e.,
$\varepsilon(r') = \alpha(r')\Delta r$. The idea is that the mass
distribution from the small spheres at each cell center should
approximate a smooth mass distribution instead of a sum of
$\delta$-functions. We optimize $\alpha(r')$ by minimizing the
relative error in $\Psi(r,\phi)$ for sharply peaked Gaussian mass
distributions located at $r'$ ranging from 0.6 to 1.8 (scaled value
for the disk-planet problem) with $c_s = 0.05$. We find a piecewise
linear function for $\alpha(r)$:
\begin{equation}
\alpha(r) = \left\{\begin{array}{ll}
0.17 + \frac{0.06}{0.6}(r-r_{\min}),  & r < 1.0 \\
0.23 + \frac{0.03}{0.2}(r-1.0),      & 1.0\le r <1.2 \\
0.26 + \frac{0.04}{0.3}(r-1.2),      & 1.2\le r <1.5 \\
0.30 + \frac{0.03}{0.2}(r-1.5),      & 1.5\le r <1.7 \\
0.33 + \frac{0.04}{0.3}(r-1.7),      & 1.7\le r
\end{array}\right.
\label{softn}
\end{equation}
Our experiments show that $\alpha(r)$ is fairly insensitive to different
resolutions in the radial direction. \cite{BaMa08}, however, have
argued that the
softening must scale with $r$. This might
be because the logarithmic grid in the radial direction is used in their FFT
implementation. For the uniform grid along the radial direction, we
find that even a constant $\alpha$, e.g., $\alpha(r)=0.23$, yields
good results for the density peak near $r=1$.

We remark that our choice of the softening should not be mixed with the
softening that is used for calculating the gravity from the planet. It has been
pointed out by \cite{NeBe03a} that the softening used between the disk and
planet should be at least 0.75 times the physical size of the grid
zone. \cite{BaMa08} suggested using $\varepsilon(r) = 0.3H(r)$ for both
disk self-gravity and the gravity between disk cells and planet. We
find that this choice is too big and produces large error for disk
self-gravity in our implementation. Therefore we use two different
softenings in our simulations: for the disk self-gravity $\varepsilon =
\alpha(r)\Delta r$, and for the gravity between disk and planet
$\varepsilon = 0.1H(r_p)$ ($r_p$ is the planet position in radial
direction).  The coefficient 0.1, which is still much smaller than the values
generally used in the literature, is chosen partially due to the
presence of vertical structure, and it matches very well with the
\cite{Tanaka02} theory on the torque evaluation. 
 
A question arises that the different softening choices may introduce
inconsistency between the disk and planet. This concern can be
alleviated by the fact that the gravity force for disk cells near the
planet are dominated by the planet gravity. Therefore this
inconsistency does not have much impact on the dynamics of the disk.   

To eliminate the dependence on the softening, we have implemented the
other approach, which is to use different grids for $r$
and $r'$ so that $r'_i$ is never equal to $r_j$. This approach has
been used by \cite{ChPO06} for their spectral method. In our
case, the integral of Eq. (\ref{2d_eq3}) is performed on a fixed
uniform grid of $r'$. Because $r'$ is the cell-centered grid, we
define $r$ as the node centered grid, i.e.,
\begin{equation}
r_0 = r'_1 - 0.5\Delta r, \qquad r_i = r'_i + 0.5\Delta r,
i=1,2,...,N.
\label{shiftG}
\end{equation}
Note that the $r$ grid has one more points than the $r'$ grid. After
calculating $\mathcal{G}(r,r',\phi-\phi')$ for $r$ at the
node-center, we can obtain the $\mathcal{G}(r,r',\phi-\phi')$ for
$r$ at the cell-center by interpolation, and then calculate the
potential $\Psi(r,\phi)$ at the cell-center. We can also calculate
directly the potential $\Psi(r,\phi)$ with $r$ at the node center
for use of the field calculation.

\subsection{Field Calculation}

We have implemented two approaches to calculate the field components. The
first approach is the difference method: we calculate the potential
first and then use the finite-difference method to calculate the
field.

If the potential is calculated at the cell-center $(r_i,\phi_j)$, we use the
central-difference to obtain the field:
\begin{eqnarray}
\Psi_r(r_i,\phi_j) &=&
\frac{\Psi(r_{i+1},\phi_j)-\Psi(r_{i-1},\phi_j)}{2\Delta r} \label{fd1}\\
\Psi_\phi(r_i,\phi_j) &=&
\frac{\Psi(r_i,\phi_{j+1})-\Psi(r_i,\phi_{j-1})}{2r_i\Delta \phi} \label{fd2}~~,
\end{eqnarray}
where the notation $(\cdot)_\phi = \partial(cdot)/(r\partial\phi)$. 
Note that we need values of $\Psi(r_0,\phi)$ and $\Psi(r_{N+1},\phi)$
for the $\Psi_r$ component at the boundary cells located at $r_1$ and $r_N$;
otherwise, one-side finite-difference, which is not very accurate,
should be used. 

If the potential is calculated at the edge-center
$(r_{i+\frac12},\phi_j)$, the field can be calculated as
\begin{eqnarray}
\Psi_r(r_i,\phi_j) &=&
\frac{\Psi(r_{i+\frac12},\phi_j)-\Psi(r_{i-\frac12},\phi_j)}{\Delta
  r}\label{fd3} \\
\Psi_\phi(r_i,\phi_j) &=&
\frac{\Psi(r_{i+\frac12},\phi_{j+1})+\Psi(r_{i-\frac12},\phi_{j+1})-
  \Psi(r_{i+\frac12},\phi_{j-1}) - \Psi(r_{i-\frac12},\phi_{j-1})
}{4r_i\Delta \phi}~. \label{fd4}
\end{eqnarray}

High order methods that use wide stencils are also possible. In fact,
we can obtain the derivative $\Psi_\phi$ with only one extra FFT in
the above algorithm of calculating the potential.
After obtaining $\hat\Psi_m(r)$, we take the inverse transformation of
$-im\hat\Psi_m(r)$ to get the $\Psi_\phi$.

The second approach for the field calculation is the integral
method: we directly convolute the field Green's function with the
mass distribution. The differentiation in $\Psi_r$ and $\Psi_\phi$
can be applied directly to the Green function, which yields
\begin{eqnarray}
\Psi_r(r,\phi) &=&
\int_{r_{\min}}^{r_{\max}}\int_0^{2\pi}\Sigma(t,r',\phi')r'
\mathcal{G}_r(r,r',\phi-\phi')dr'd\phi'
\label{Gr1}\\
\Psi_\phi(r,\phi) &=&
\int_{r_{\min}}^{r_{\max}}\int_0^{2\pi}\Sigma(t,r',\phi')r'
\mathcal{G}_\phi(r,r',\phi-\phi')dr'd\phi' \label{Gphi1}
\end{eqnarray}

Since we have two field components, $\Psi_r$ and $\Psi_\phi$, the
integral method takes twice the computation time of the difference
method. But, we expect that the integral method might have higher accuracy
than the finite difference approximation. However, if the
mode cut-off approach, which will be described in Section
\ref{sec:para}, is used, it is possible that the integral method may
lose this advantage. In 
fact, we observed in our numerical tests that the integral method of
(\ref{Gr1}) and (\ref{Gphi1})
with the mode cut-off gave larger errors and more than
double the computation and communication times of the
finite-difference method.

\subsection{Tree-code force calculation\label{sec:tc}}

Yet another approach we explored to solve for the field involved
using a hierarchical tree-code.  In this approach, we treat each cell
as a particle, and hence an effective
force law between two points, $(r,\phi)$ and $(r',\phi')$, in the
disk that accounts for the vertical structure is required. Just as
before, we assume $\rho({\bf x}) = \Sigma(r,\phi)Z(r,z)$. And since
symmetry arguments require $Z(r,z)$ to be an even function of $z$,
the z-component of the force must be zero at the $z=0$
plane. Therefore, the force is 
directed in the plane of the disk and it has a magnitude factor given by
\begin{equation}
F(d) = \int_{-\infty}^{\infty}\int_{-\infty}^{\infty}
\frac{Z(r,z)Z(r',z')d}{\big(d^2 +
(z-z')^2\big)^{3/2}}dzdz'\label{eq_tc1}
\end{equation}
where $d = \big(r^2+{r'}^2 - 2rr'cos(\phi-\phi')\big)^{1/2}$.

If we further assume $Z(r,z)$ is given by Eq. (\ref{eq_z}), then we
can rewrite Eq. (\ref{eq_tc1}) as
\begin{equation}
F(d;H,H') = \frac{C_{3D}(d;H,H')}{d^2} \label{eq_tc2}
\end{equation}
where
\begin{equation}
C_{3D}(d;H,H') = \frac{d^3}{2 \pi H
H'}\int_{-\infty}^{\infty}\int_{-\infty}^{\infty} \frac{e^{-z^2/2
H^2} e^{-z'^2/2 H'^2}}{\big(d^2 + (z-z')^2\big)^{3/2}}dzdz'
\label{eq_tc3}
\end{equation}
is called the 3D correction factor with respect to the 2D force
without the vertical structure. As $H\to 0$, (\ref{eq_tc3})
approaches to the 3D factor defined by \cite{Koller04}.

We note that this force law is a line-line force (giving the net
interaction between the mass distributed along two vertical lines at
$(r,\phi)$ and $(r',\phi')$). This is different from what is done in
the Green's function method, which uses a point-line interaction. By
setting $H=0$ in Eq. (\ref{eq_tc2}), we can recover the interaction
force law of the Green's function method.  However, if we wished to
include line-line interactions in the Green's function method, we
would have to abandon the more efficient potential calculation of
Eq. (\ref{intGz}) and instead use Eqs. (\ref{Gr1}) and
(\ref{Gphi1}) with $\mathcal{G}_r$ and $\mathcal{G}_{\phi}$ defined
appropriately.

Re-scaling the integration in Eq. (\ref{eq_tc3}) and introducing the
non-dimensional variables $d' = \frac{d}{\sqrt{H^2 + H'^2}}$ and
$\alpha = \frac{H'}{H}$ gives
\begin{equation}
C_{3D}(d';\alpha) = \frac{d'^3}{2
\pi}\int_{-\infty}^{\infty}\int_{-\infty}^{\infty} \frac{e^{-z^2/2}
e^{-z'^2/2}}{\bigg(d'^2 + \frac{(z-z'\alpha)^2}{1 +
\alpha^2}\bigg)^{3/2}}dzdz' \label{eq_tc4}
\end{equation}
Finally, we perform an orthogonal coordinate transformation using
the variables
\begin{eqnarray}
\zeta = \frac{1}{\sqrt{1+\alpha^2}}(z - z'\alpha) \nonumber
\\
\zeta' = \frac{1}{\sqrt{1+\alpha^2}}(\alpha z + z'). \nonumber
\end{eqnarray}
This allows us to simplify Eq. (\ref{eq_tc4}) to
\begin{equation}
C_{3D}(d') = \frac{d'^3}{2
\pi}\int_{-\infty}^{\infty}\int_{-\infty}^{\infty}
\frac{e^{-\zeta^2/2} e^{-\zeta'^2/2}}{\big(d'^2 +
\zeta^2\big)^{3/2}}d\zeta d\zeta' \label{eq_tc5}
\end{equation}
revealing that it is actually independent of $\alpha$.  In fact,
similar to Eq. (\ref{intGz}) we can express Eq. (\ref{eq_tc5}) as
\begin{equation}
C_{3D}(d') = -\frac{d'^3}{2 \sqrt{2 \pi}}e^{d'^2/4}\bigg(K_0(d'^2/4)
+ K'_0(d'^2/4)\bigg) \label{eq_tc6}
\end{equation}
where $K_0$ and $K'_0$ denote the modified Bessel function of the
second kind and its derivative.

The tree-code requires the calculation of $C_{3D}(d')$ and its first
and second derivatives over a wide range of $d'$ (ranging from
$10^{-2}$ to $10^3$).  Calculating Eq. (\ref{eq_tc6}) on the fly is
prohibitively expensive and should be avoided.  Another option is to
pre-compute Eq. (\ref{eq_tc6}) and its derivatives at a discrete set
of $d'$ values. Then when $C_{3D}(d')$ is needed, we interpolate the
pre-computed data.  This approach can provide a significant speedup,
but because the data is stored in main memory it is still too slow
for our purposes.

\begin{figure}[htbp]
\begin{center}
\includegraphics[width=0.45\textwidth]{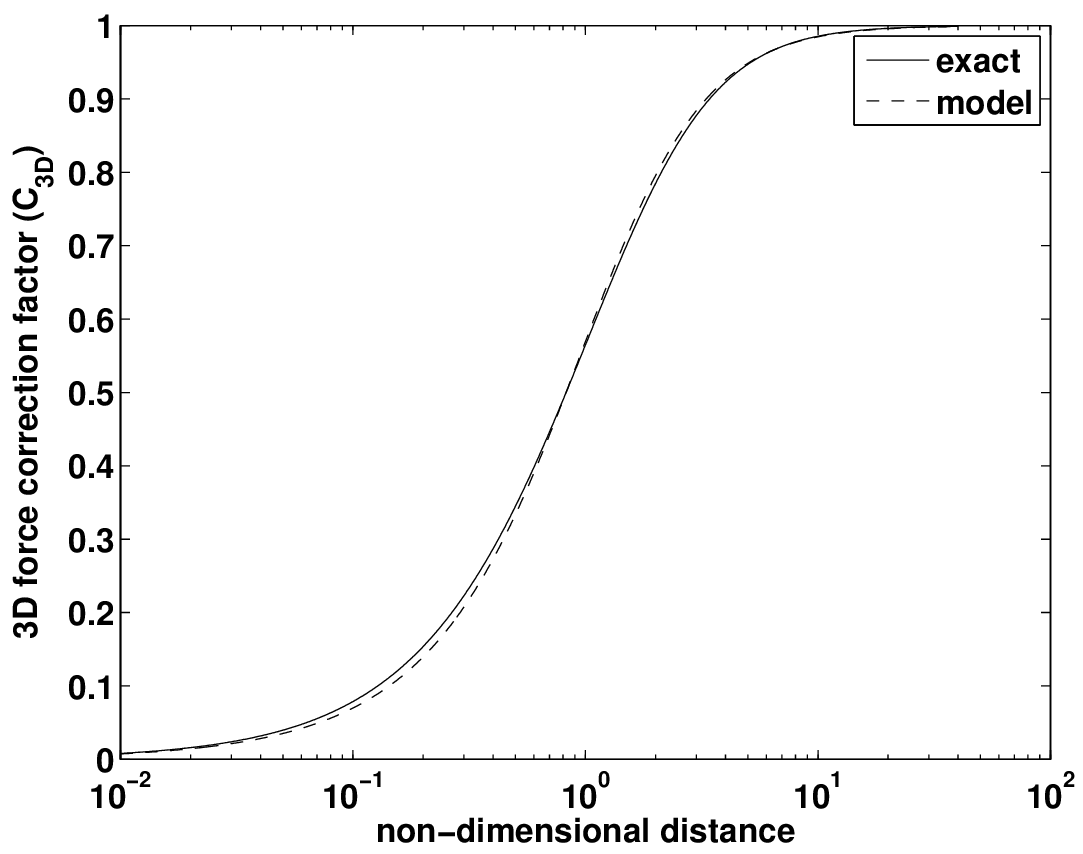}
\includegraphics[width=0.45\textwidth]{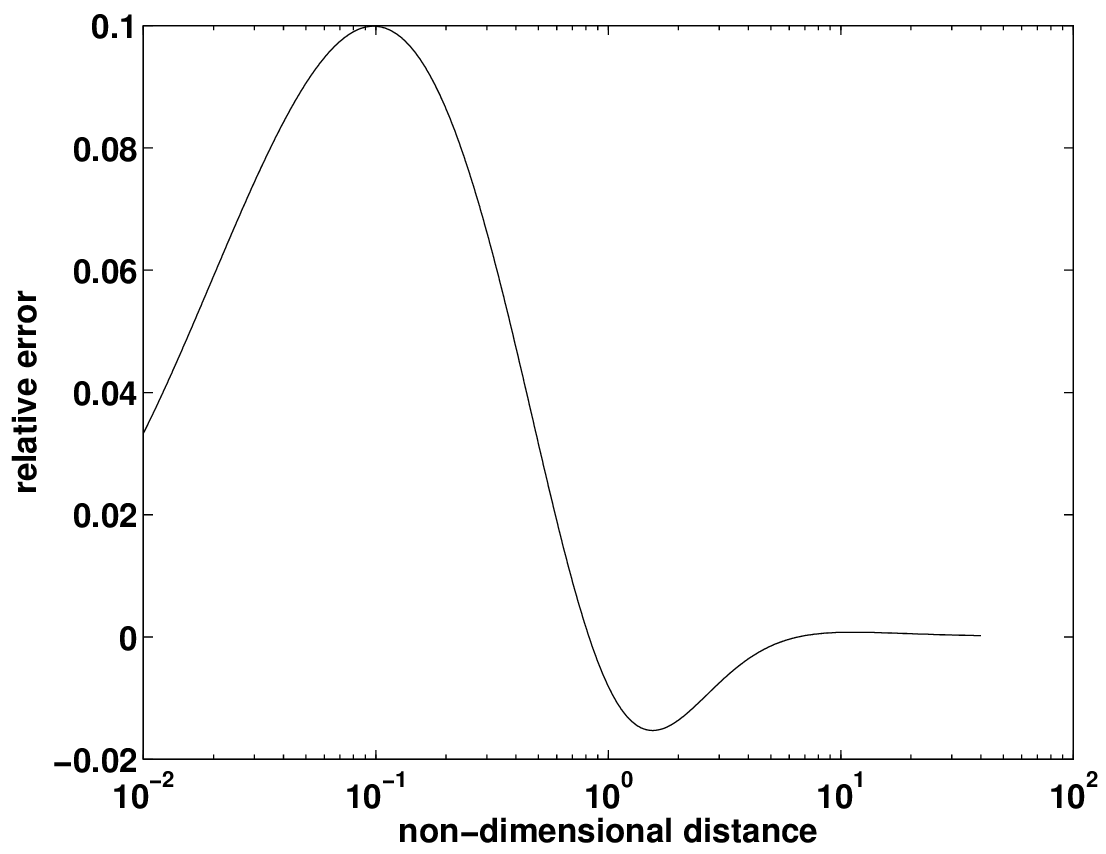}
\caption{\label{fig_tc1} 3D correction factor (Eq. (\ref{eq_tc6})) and
  relative error
plotted vs. non-dimensional distance ($d'$) using model Eq. (\ref{eq_tc7}).}
\end{center}
\end{figure}
\begin{figure}[htbp]
\begin{center}
\includegraphics[width=0.45\textwidth]{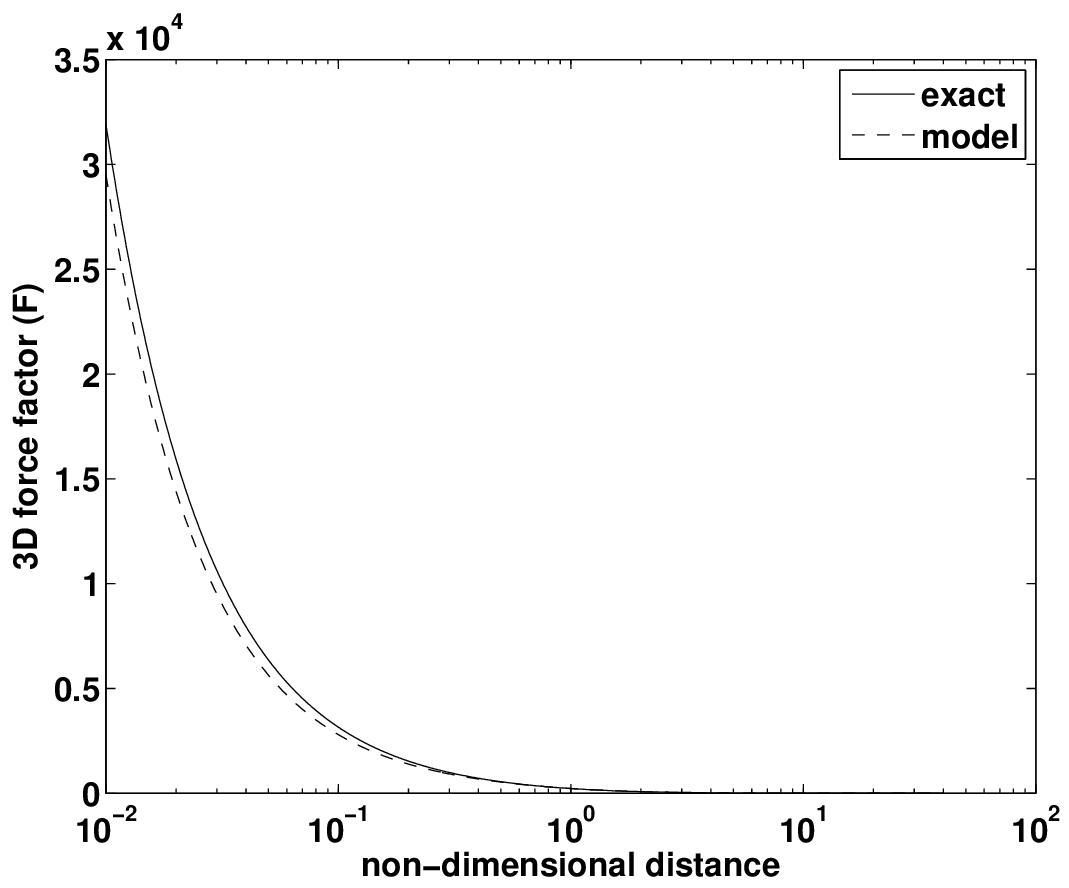}
\includegraphics[width=0.45\textwidth]{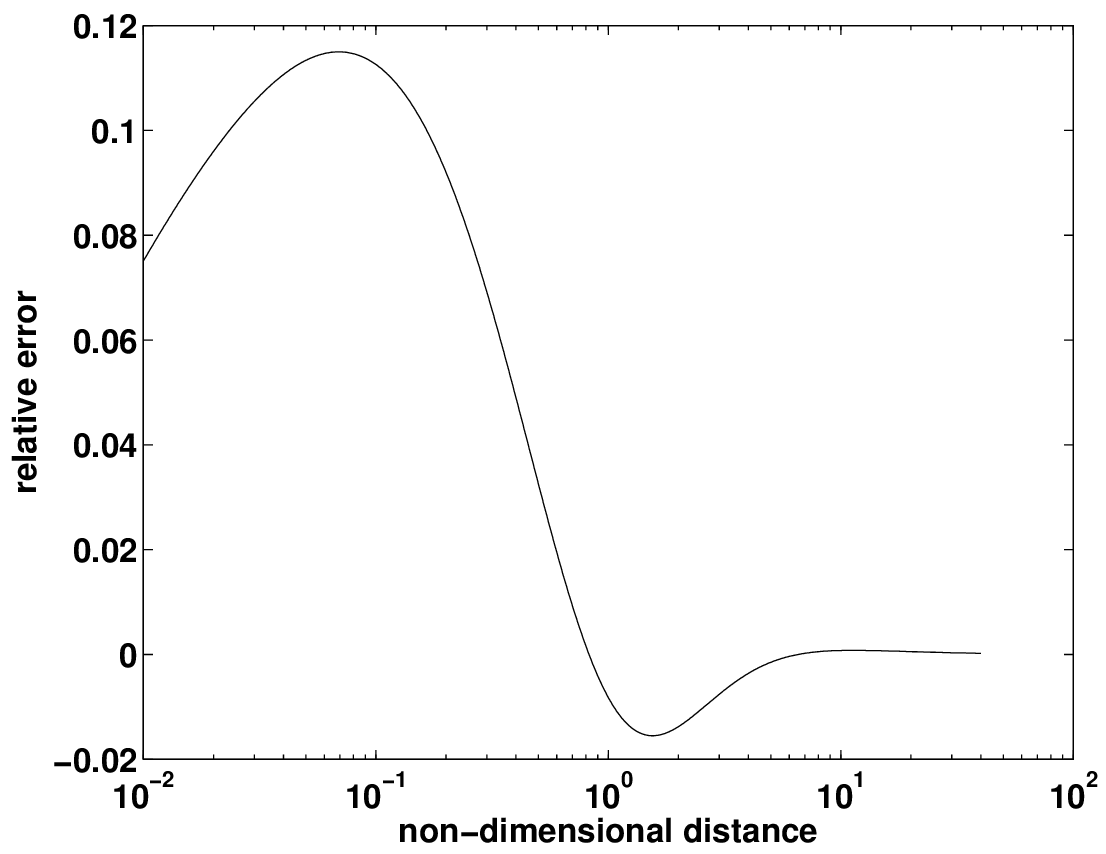}
\caption{\label{fig_tc2} 3D Force factor (Eq. (\ref{eq_tc2}))
  and relative error plotted vs. non-dimensional distance ($d'$) using model
  Eq. (\ref{eq_tc7}).} 
\end{center}
\end{figure}

Another approach we explored involves approximating Eq.
(\ref{eq_tc6}) with a model function that is accurate yet
inexpensive enough to compute on the fly.  Plotting $C_{3D}(d')$
reveals its form (Fig. \ref{fig_tc1}) and suggests a model function
of the form
\begin{equation}
C_{3D}^{(\mathrm{model})}(d') = \frac{(d'/d'_{1/2})^p +
(d'/d'_{1/2})^{2p}}{2 + (d'/d'_{1/2})^p + (d'/d'_{1/2})^{2p}}.
\label{eq_tc7}
\end{equation}
We find optimal parameter values to be $d'_{1/2} = 0.8252$ and $p =
0.957$.  The relative error introduced by using this model is plotted with
$C_{3D}$ and $F$ in Fig. \ref{fig_tc1} and 
\ref{fig_tc2}. Since generally $d' = d/H(r) > \varepsilon/H(r)$ with $H'=0$,
the model is acceptable for distance or softening larger than
$0.005H(r)$. 
This model requires the calculation of one power
function, a few additions and multiplications and one division. For
our application, we found this to be the most efficient way to
include 3D structure using a tree-code.

Our tree-code is implemented in a distributed memory parallel
architecture using MPI, just like our Green's function FFT
algorithm.  There are many papers detailing parallel tree-codes
\citep{Dubin96,MiDo02}. So we simply highlight the important points of our
algorithm here.

\begin{enumerate}
\item We begin by pre-computing the tree structure of our grid,
which does not change with time.  Each processor retains a copy of
this {\it quad-tree} to be used throughout the simulation.
\item All processors compute partial sums (using local particles) of the
moments for all nodes in the tree.
\item These partial sums are added and broadcast back to all the
processors.
\item A complete copy of the density profile is distributed to all
the processors.  This is required for load
balancing and computing the direct summation part of the force (step
5(c)).
\item Forces are computed at every grid cell. The disk is partitioned
azimuthally (the symmetry providing ideal load balancing), and
processors are assigned equal-sized sectors of particles (grid
cells). Each particle traverses the {\it quad-tree} to include
interactions with all other particles.
    \begin{enumerate}
    \item If a particle-node interaction meets the accuracy
    criterion, then the particle interacts with that node via the
    moments computed in step 3.
    \item If the accuracy criterion is not met (the node is too large or
    the particle-node separation too small), then we drop one level
    in the tree and check all the rejected node's sub-nodes for
    particle-node interactions.
    \item Last, if we reach a leaf-node (a node with no sub-nodes) and the
    accuracy criterion is not met we perform direct summation over all the
    leaf-node's particles.
    \end{enumerate}
\end{enumerate}

\section{Numerical Implementation and Experiments\label{sec:test}}

\subsection{Description of the test problem}

As in \cite{ChPO06}, we consider the density function to be a
Gaussian sphere given by
$$
\rho(r,\phi,z) =
\frac{M}{(2\pi\sigma^2)^{3/2}}\exp\left(-\frac{r^2+z^2}{2\sigma^2}\right),
$$
where $M$ is the normalized total mass, $\sigma$ is a parameter that
controls the width of the mass distribution and it has similar role as the
scale height in Eq. (\ref{eq_z}).
The potential on the $z=0$ plane has been given by \cite{ChPO06}
\begin{equation}
\psi(r,\phi) = -\frac1r\mbox{erf}\left(\frac{r}{\sqrt{2}\sigma}\right)~,
\label{eq_psi}
\end{equation}
where $\mbox{erf}(x)$ is the error function
$$
\mbox{erf}(x) = \frac{2}{\sqrt{\pi}}\int_0^xe^{-t^2}dt ~~.
$$
For a collection of Gaussian spheres centered on the $z=0$ plane with
the same $\sigma$, the $z$-dependent vertical structure can be
factored out, i.e.,
$$
\rho(r,\phi,z) = \sum_i\rho_{(r_i,\phi_i)}(r,\phi,z) =
Z(r,z)\sum_i\Sigma_{(r_i,\phi_i)}(r,\phi),
$$
where $(r_i,\phi_i)$ is the center of each sphere, $Z(r,z)$ gives
the vertical structure
\begin{equation}
Z(r,z) = \frac{1}{\sqrt{2\pi
    \sigma^2}}\exp\left(-\frac{z^2}{2\sigma^2}\right),
\label{eq_z2}
\end{equation} 
$\Sigma_{(r_i,\phi_i)}$ is the surface density
$$
\Sigma_{(r_i,\phi_i)} =
\frac1{2\pi\sigma^2}\exp\left(-\frac{R_i^2}{2\sigma^2}\right),
$$
and $R_i(r,\phi) = \sqrt{r^2 +{r_i}^2 - 2rr'\cos(\phi-\phi_i)}$ is
the distance between $(r,\phi)$ and $(r_i,\phi_i)$.  Notice that Eq.
(\ref{eq_z2}) is the same as Eq. (\ref{eq_z}) with a constant scale
height $H(r)=\sigma$.

We use three Gaussian spheres located at $(r,\phi)=(1,0)$,
$(0.9,3\pi/4)$, and $(1,-\pi/2)$ with a total mass of 2,1/2, and 1
respectively, i.e.,
\begin{equation}
\Sigma_{\mbox{ana}}(r,\phi) = 2\Sigma_{(1,0)}(r,\phi) +
\frac12\Sigma_{(0.9,3\pi/4)}(r,\phi) + \Sigma_{(1,-\pi/2)}(r,\phi),
\label{den_exact}
\end{equation}
where $(*)_{\mbox{ana}}$ represents the analytic solution.
The exact potential will be
\begin{equation}
\Psi_{\mbox{ana}}(r,\phi) = 2\psi(R_{(1,0)}) + \frac12\psi(R_{(0.9,3\pi/4)}) +
\psi(R_{(1,-\pi/2)})
\label{pot_exact}
\end{equation}
where $\psi$ is defined by Eq. (\ref{eq_psi}). The exact field
$\Psi_r$ and $\Psi_\phi$ can also be calculated accordingly.

The tests are done in the computational domain
$[0.4,2.0]\times[0,2\pi]$ with grid $N_r\times N_\phi$. The normalized
total mass for the disk is $M=0.002M_\star$. We choose
$N_\phi=4N_r$ so that the cells near $r=1$ are nearly square. Without
specification, we always use $N_r=800$, which reaches the convergence
in both the torque calculation on planet and the azimuthal averaged potential
vorticity distribution on the disk in our disk-planet interaction
simulations \citep{lietal05}. 
Since the surface density (\ref{den_exact}) approaches to zero outside the
computational 
domain, the exact potential (\ref{pot_exact}) is still valid for our
truncated domain. For
convenience, we use the following notations for comparison: $N_p$ denotes
the number of 
processors, $E_{\max}$ denotes the maximum error over the whole
domain, $RE_{\max}$ denotes the maximum relative error, and $p$ is the
convergence order in $E_{\max}$. Since the force is a vector that has
two component, we use the following formula to calculate $E_{\max}$
$$
E_{\max} = \sqrt{|f^r - f^r_{\mbox{ana}}|^2 + |f^\phi - f^\phi_{\mbox{ana}}|^2}
$$
where $f^r$ and $f^\phi$ are the force components in $r$- and
$\phi$-direction respectively. 
The local relative error for a variable $u$ is defined as  $RE_{\max}(u) =
{E_{\max}}/{|u_{\mbox{ana}}|}$. For the force,
$|u_{\mbox{ana}}|=\sqrt{(f^r_{\mbox{ana}})^2 + (f^\phi_{\mbox{ana}})^2}$.
The global relative error for a variable $u$ is defined as 
\begin{equation}
RE(u) =
\frac{\sum_{j=1}^{N_\phi}\sum_{i=i}^{N_r}|u_{i,j}-u_{i,j,\mbox{ana}}|}
{\sum_{j=1}^{N_\phi}\sum_{i=i}^{N_r}|u_{i,j,\mbox{ana}}|}
\label{eq_err_re}
\end{equation}

All of the computation are performed on a parallel Linux cluster at
the Los Alamos 
National Lab. Each node of the cluster is dual-core AMD  Opteron(tm)
processor with 2.8G HZ and 2GB local memory.   

\subsection{Parallel implementation and comparison\label{sec:para}}

Our hydro simulation for the interaction of disk and proto-planet
problem is performed on a high resolution grid, e.g., $N_r\times
N_\phi=800\times3200$ grid. 
The whole domain is split into annular regions for
parallel computation. Each annular region has the same number of
cells in the radial direction.

The Fourier transform and its inverse can be parallelized without
much difficulty because they require no communication between
different processors. However, the numerical quadrature of Eq.
(\ref{2d_eq3}) is done in the radial direction, where each processor
holds only part of the information about the density distribution in
the whole domain. 

We have tested two approaches to parallelize the numerical
quadrature. In the first approach, we start by computing a partial
quadrature including only source terms from the local density
distribution, i.e. each processor computes a partial sum of Eq.
(\ref{2d_eq3}) for all $r$ and $m$. We then apply a global
communication to obtain the complete summation. 

In the second approach, we begin by performing a global
communication over the whole domain so that every processor has a
complete copy of the density profile for
$r'\in[r_{\min},r_{\max}]$. Then the complete numerical 
quadrature (\ref{2d_eq3}) can be performed to obtain
$\hat\Psi_m(r)$ for the local portion of $r$ grid. 

We remark that in the pre-computing stage, no matter which approach is
used, each 
processor can compute its own portion of the  
Fourier transform $\hat{I}_m(r,r')$ independently and store it for
the later use. It does not need to have a copy of whole profile for
all $r$ and $r'$. 

In the following we will first propose a strategy to reduce the
parallel communication cost, and then compare the above two approaches in
calculating potential $\Psi(r,\phi)$. For simplicity, we denote the
first approach as approach I, and the second as approach II. 

\subsubsection{Mode cut-off approach to reduce the communication cost}

The communication cost in both approaches is proportional to both
the number of processors and the amount of data being communicated
(total grid size). Consequently, we expect that the communication cost
will dominate the computation cost as the number of processors is
increased, eventually becoming a bottleneck for a large number of
processors. Note that the communication is done in the
frequency domain, where the high modes decay exponentially for a
smooth function. Thus we can truncate the Fourier modes above a cutoff
parameter ($M_{cut}$) without a significant loss of accuracy. If
$M_{cut}$ is far smaller than $N_\phi$ (the total number of mode), the
communicate cost can be 
reduced to a small of fraction of the cost before the cutoff.

For both approaches, we can calculate $M_{cut}$ based on the Fourier
transformation of the density profile. For a specific radius $r_i$, we
define the energy contained in $\hat\Sigma(r_i)$ as
$\sum_{m=0}^{N_\phi}|\hat\Sigma_m(r_i)|^2$. Each processor then computes
an $m_{cut}(r)$ based on satisfying some preset fraction
$\epsilon_{cut}$ (e.g., $\epsilon_{cut}=10^{-4}$) of the total
energy for modes above $m_{cut}(r)$. Next, each processor finds its
maximum $m_{cut}$ over its range of $r$.  Last, all the processors
compare their $m_{cut}$ values to obtain the global maximum
$M_{cut}$ that is used to truncate the Fourier series. Note that
$M_{cut}$ calculated in this way does not vary with the number of
processors.

For approach I, the $m_{cut}(r)$ can also be evaluated based on
$\hat\Psi(r)$ instead of $\hat\Sigma(r)$, because the communication is
done after the partial summation. 
To be more accurate in the field calculation, we instead can evaluate
$m_{cut}(r)$ based on the 
energy defined by $\sum_{m=0}^{N_\phi}|m\hat\Psi_m(r_i)|^2$. The extra
factor of $m$ is included so that we obtain the energy of the force,
which is the derivative of $\Psi$. Notice, however, that if we
compute $m_{cut}$ and $M_{cut}$ as described above, $M_{cut}$ will
vary with the number of the processors. This is because each
processor computes only a partial sum of $\hat\Psi_m(r)$. To produce
a relatively constant $M_{cut}$, after each processor finds its
maximum $m_{cut}$, all the processors take an energy-weighted
average of their $m_{cut}$ to obtain the global $M_{cut}$.  It is
observed that the energy-weighting average gives an $M_{cut}$ that
remains nearly constant while the number of processors varies
between 1 to $N_p$. For verification reason, we can require that
$M_{cut}$ be independent of the number of processors. However, that
requires a costly global communication between different processors.
Nonetheless, we find that $M_{cut}$ will only increase slightly 
with the number of processors, which means that increasing the
number of processors will not degrade the numerical accuracy. We
define a similar preset fraction $\epsilon_{cut}$ to the previous
density mode cut-off approach. After extensive experiments, we
observe that $\epsilon_{cut}\approx 10^{-7}$ in $\hat\Psi(r)$
cut-off produces similar results to $\epsilon_{cut} = 10^{-4}$ in
$\hat\Sigma(r)$ cut-off.

For density distributions in disk-planet simulations, the energy of the
zero mode ($m=0$) often dominates that of all other modes
combined. Thus, when calculating the energy using $\hat\Sigma$ or
$\hat\Psi$ , it is convenient to exclude the zero mode from the 
total energy. Notice that if the energy is
defined as $|m\hat\Psi_m(r)|^2$, the zero-mode is naturally
excluded. To remove the numerical noise, we multiply the energy of
the zero mode by a small number (e.g., $\epsilon$) and add it to the
total energy. In our disk-planet simulations, the initial density
distribution is axisymmetric, and hence zero mode is enough. With the
time evolution of the surface density, $M_{cut}$ begins to increase
quickly until it reaches a certain number, which is insensitive
to different resolutions and different initial power-law density profiles. 
However $M_{cut}$ does vary a lot with the planet mass and sound
speed. For an example, we obtain the following data for simulations of
an isothermal disk with different planet mass ($M_p$) and sound speed ($c_s$), 
$$
M_{cut}=\left\{\begin{array}{cl}
  190, & \mbox{    for } M_p=3\times10^{-5}M_\star, c_s = 0.05 \\
  297, & \mbox{    for } M_p=3\times10^{-4}M_\star, c_s = 0.05 \\
  298, & \mbox{    for } M_p=3\times10^{-5}M_\star, c_s = 0.035 
\end{array}\right.
$$
Note that $M_{cut}$ is much smaller than the total number of modes in
$\phi$-direction, $N_\phi=3200$. 

\subsubsection{Impact of the cutoff modes and cut-off
  thresholds\label{sec:test_cut-off}} 

Here, we study how the number of cut-off modes, $M_{cut}$, impacts
the computational efficiency and accuracy. We test the problem
with parallel computation using 100 processors. 
We will show the
results only for approach II. Similar results have also been
obtained for approach I.

\begin{figure}[htbp]
\begin{center}
\includegraphics[width=0.45\textwidth]{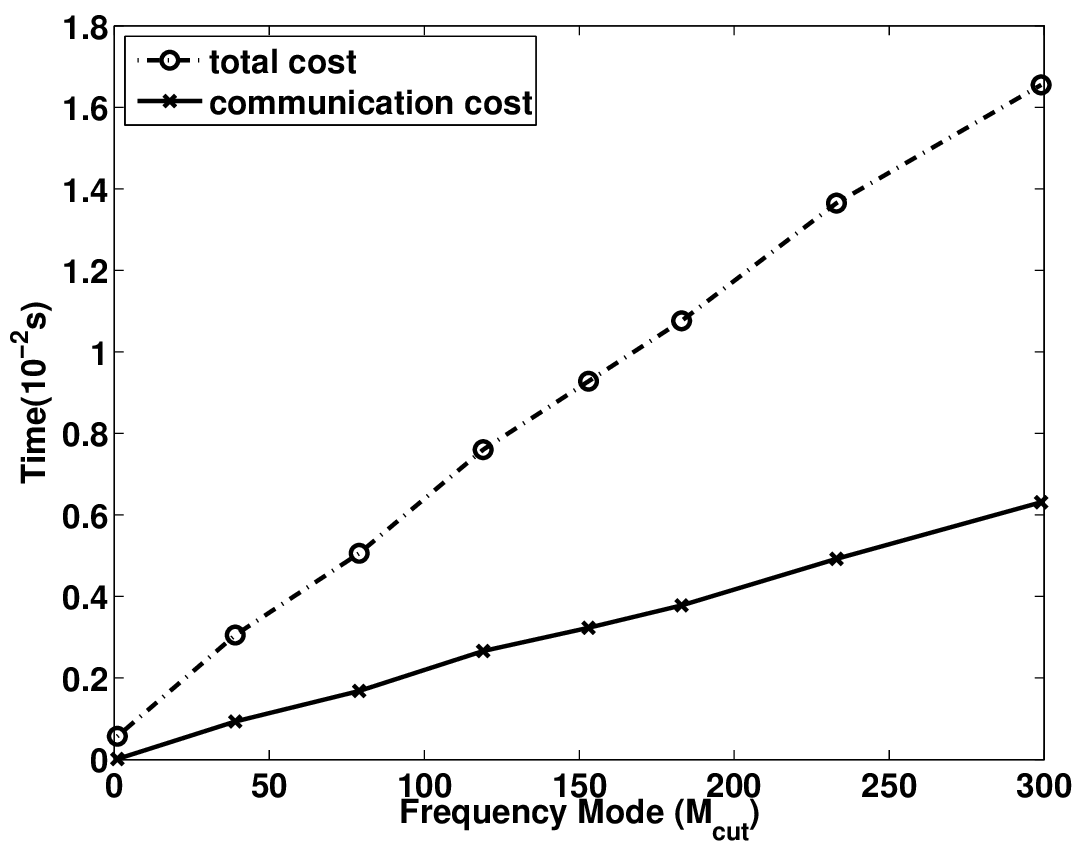}
\includegraphics[width=0.45\textwidth]{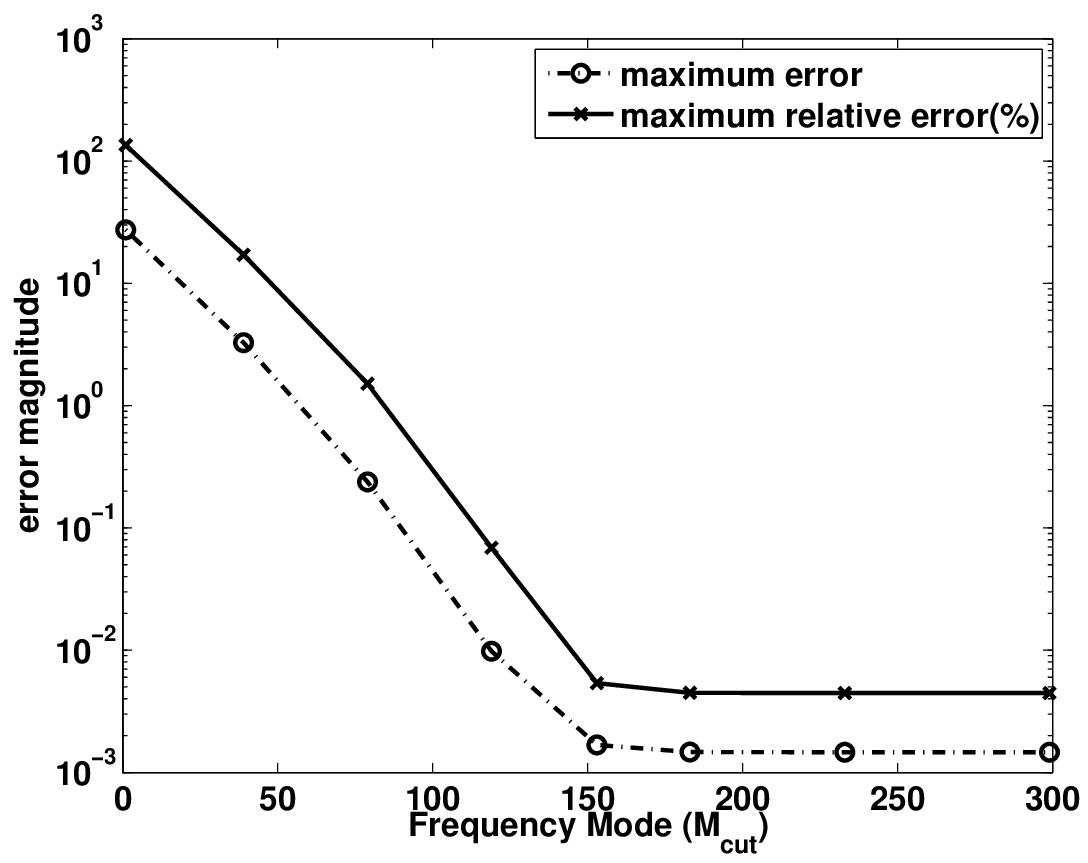}
\caption{\label{cpu1}The computation and communication cost for
  evaluating $\hat\Psi(r,\phi)$ (left) and the error variation 
  (right) as a function of $M_{cut}$. 
}
\end{center}
\end{figure}

Fig. \ref{cpu1} shows the variation in the computation and
communication time with different numbers of cut-off modes. The
total cost displayed in term of CPU time includes the computation of
one FFT of the 
density field, the global communication of the density spectrum up-to 
the cut-off mode $M_{cut}$ right after the FFT, 
a quadrature calculation in the radial direction. Fig. 
\ref{cpu1} shows that the communication time
increases at a slower rate with $M_{cut}$ than the computation time .

\begin{figure}[htbp]
\begin{center}
\includegraphics[width=0.45\textwidth]{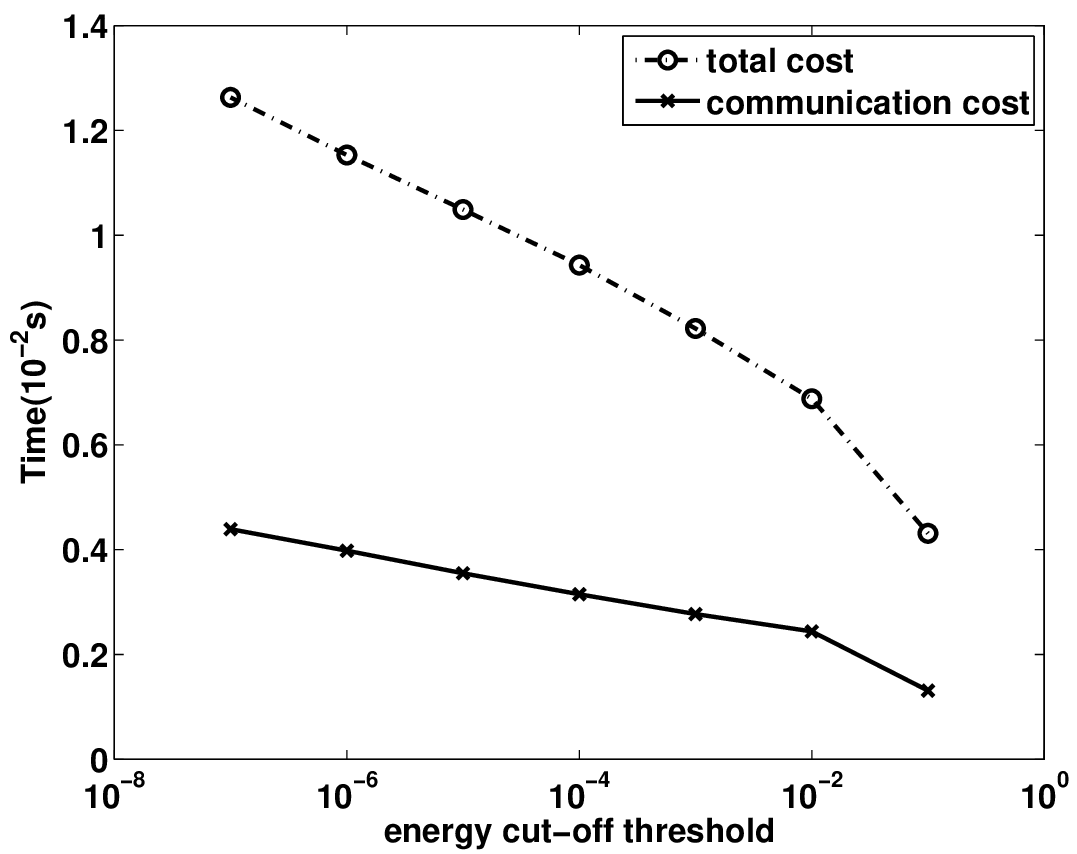}
\includegraphics[width=0.45\textwidth]{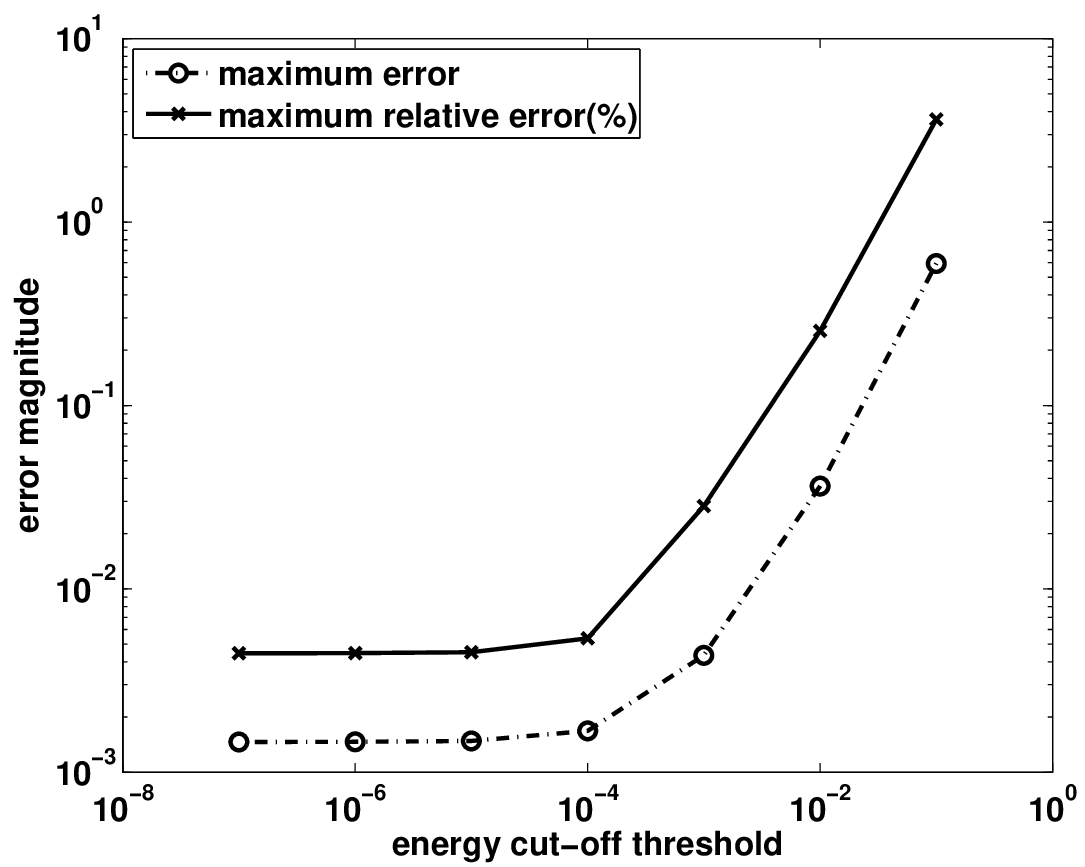}
\caption{\label{cpu2}The computation and communication cost for
  evaluating $\hat\Psi(r,\phi)$ (left) and the error variation
  (right) as a function of energy cut-off thresholds
  $\epsilon_{cut}$. The cut-off is 
  based on the energy of $\hat\Sigma$. 
}
\end{center}
\end{figure}

The energy cut-off approach depends on the cut-off threshold
$\epsilon_{cut}$. We also vary the 
size of $\epsilon_{cut}$ to see how it impacts the cut-off mode.
Fig. \ref{cpu2} shows the simulation results for
different $\epsilon_{cut}$. Figs \ref{cpu1} and \ref{cpu2} show that
our cut-off threshold, $\epsilon_{cut}=10^{-4}$, which corresponds to
$M_{cut}=153$, appears to be the optimal number
to balance accuracy and efficiency for this problem. This energy
cut-off threshold is insensitive to grids with different resolutions.

\subsubsection{Parallelization via domain decomposition of the source
  grid:  approach I}

The approach I is implemented via domain decomposition of the
source grid $r'$. In terms of the communication
cost, it involves a global reduction of $N_rM_{cut}$
data from every processor and a distribution of $N_rM_{cut}/N_p$ data
to each processor, where $N_p$ is the number of processors. 
Here we use a fixed mode cut-off number $M_{cut}=153$. 

\begin{figure}[htbp]
\begin{center}
\includegraphics[width=0.8\textwidth]{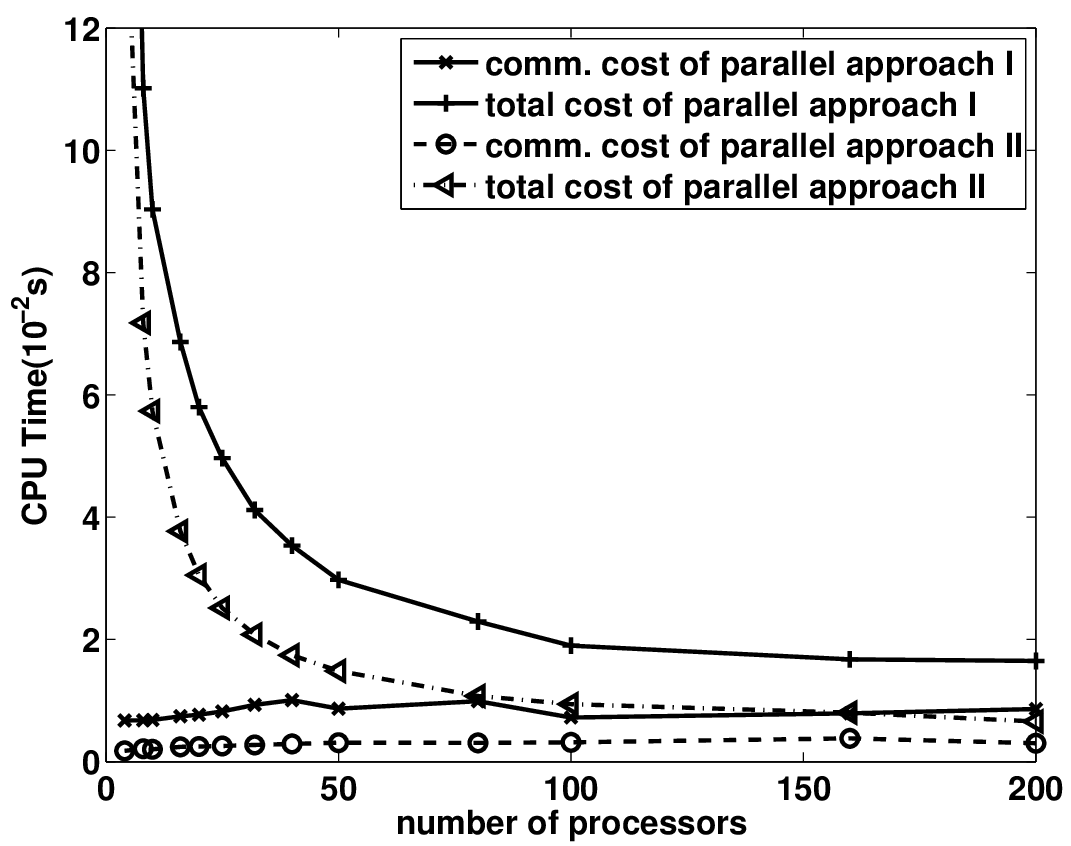}
\caption{\label{cpu3}The communication and total cost for evaluating
  $\hat\Psi(r,\phi)$ varying with the number of
  processors in MPI. $M_{cut} = 153$.
}
\end{center}
\end{figure}

Fig. \ref{cpu3} shows how the communication and computation cost vary
with the 
number processors. Note that the communication cost remains relatively
constant no matter how many number of the processors we use. This is a
surprise, because we expect that the communication cost is
proportional to the length of the
communicated data, which is proportional to the number of processors. 
We remark that one might obtain a different performance for a different MPI
implementation or a different parallel cluster. In year 2007, we
observed using the same code that the communication cost increases
linearly with the number of 
processors. In year 2008, our
parallel cluster has been upgraded with a 
new parallel software and a new inter-connection. The data shown in
Fig. \ref{cpu3} is calculated on the new cluster.

\subsubsection{Parallelization via domain decomposition of the field
  grid: approach II}

The approach II described above is implemented via domain
decomposition of the 
field grid $r$. It involves a global gathering of $N_rM_{cut}/N_p$ data from
every processor and a broadcast of $N_rM_{cut}$ data to each
processor. While the amount of communicated data is the same for both
approaches, approach I also involves a global summation operation resulting
in a total communication cost of about 10\% more than approach II with
the same value $M_{cut}$. 
Again we show some results using only a fixed cut-off mode number
$M_{cut} = 153$.   

Fig.\ref{cpu3} (approach II) shows the performance comparison with the first
approach. Similar to the results of approach I
described in the last subsection, the communication cost remains
nearly constant. It is clear that approach II (this approach) is faster than
the approach I. Therefore we will use approach II as our choice of the
methods in the tests hereafter. 

\begin{figure}[htbp]
\begin{center}
\includegraphics[width=0.8\textwidth]{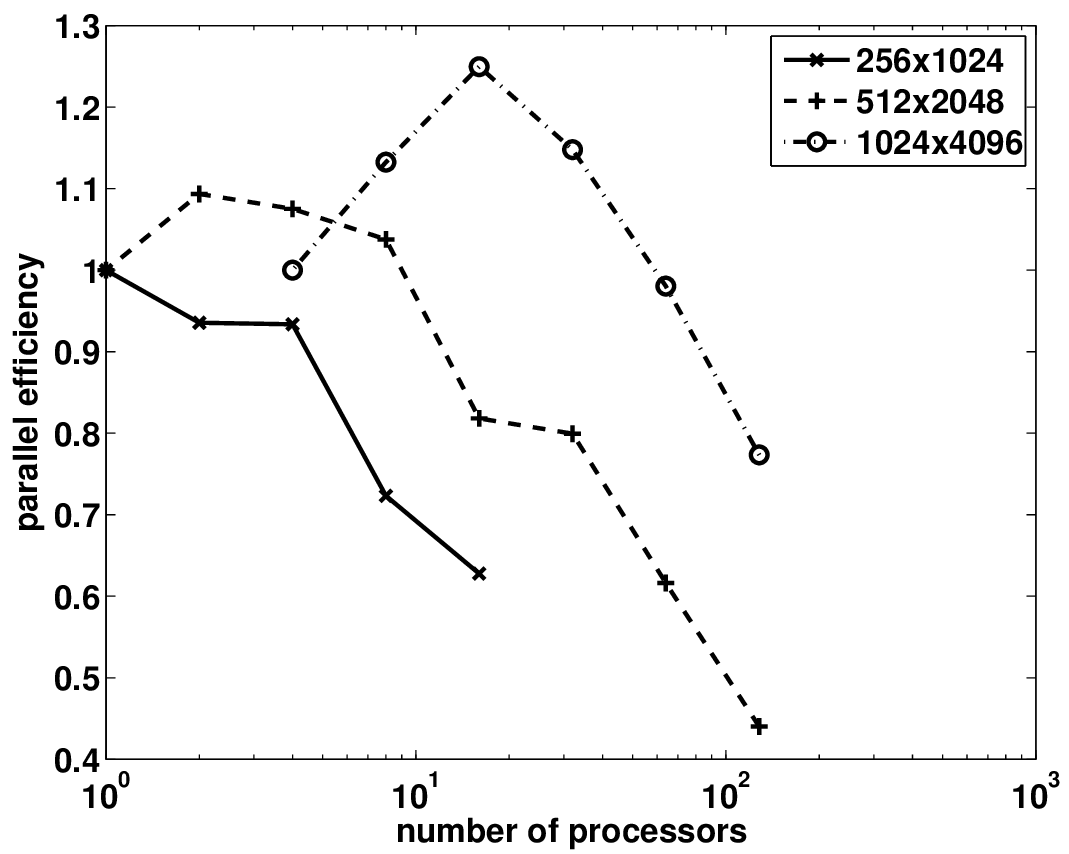}
\caption{\label{cpu4}The parallel efficiency for grids with different
  resolutions and different number of processors.
  $M_{cut} = 153$ is used.
}
\end{center}
\end{figure}

Fig. \ref{cpu4} shows the parallel
efficiency for different numbers of processors ($N_p=2^n,0\le n\le7$)
and grids with different resolutions.
The
parallel efficiency is defined by
$$
E(N_p) = \frac{T_{seq}(1)}{N_pT(N_p)}
$$
where $T(N_p)$ is the run time of the parallel algorithm, and
$T_{seq}(1)$ is the runtime of the sequential algorithm using one
processor. For the fine resolution grid $1024\times4096$, we
replace $T_{seq}(1)$ with $4T(4)$ due to the memory limitation of the
processors. 
It is clear that for a
smaller number of processors and a larger grid, the parallel
efficiency is better. Fig. \ref{cpu4} shows
that the parallel efficiency can be larger than 1 at certain
stages. The reason is not clear to us.
 
\subsection{Comparison of different approaches for potential
  calculation\label{sec:test1}} 

We have tested two approaches for computing the potential and field:
one is with the softening Eq. (\ref{softn}) and the other is without
softening but with a shifted grid defined by Eq. (\ref{shiftG}).
Note that the field is calculated differently using different central-difference 
methods. The field force with softening is calculated using
central-difference Eqs. (\ref{fd1}) and (\ref{fd2}). 
The field force without softening is
calculated using central-difference Eqs. (\ref{fd3}) and
(\ref{fd4}). 

To minimize the impact of the cut-off mode, we use a fixed cut-off
number $M_{cut}=153$, which produces nearly the same results as
without cut-off (see \S \ref{sec:test_cut-off}). 
Table \ref{tab1} shows the numerical errors for both approaches. The
convergence order is calculated based on the maximum absolute
error. The results show nearly second-order convergence as  
the mesh is refined. This is in agreement with the accuracy of our
method, because both the quadrature rule to calculate the potential and
the finite-difference method to calculate the force are of second-order
accuracy. 

Based on global relative errors ($RE$ in Table \ref{tab1}) for both
potential and field, we see that  
the potential method using the shifted grid without softening
is more accurate than with softening. Yet the difference is relatively
small. 

\begin{table}[htbp]
\begin{center}
\caption{\label{tab1} Numerical results for the potential calculation
with and without softening Eq. (\ref{softn}). The top half is for with
softening and the bottom half is for shifted grid without softening.}
\begin{tabular}{c|ccc|ccc}
       & \multicolumn{3}{c|}{Force error}  &
     \multicolumn{3}{c}{Potential error} \\
Grid & $E_{\max}$ & $RE$ & $p$ & $E_{\max}$ & $RE$ & $p$ \\
$N_r$  &        &($10^{-2}$)(1)& &($10^{-2}$)&($10^{-5}$)& \\
\hline
128 & 2.6514 & 3.1388 &    &3.2631&7.2648&     \\
256 & 0.7922 & 0.9205 &1.74&1.7595&3.0652&0.89 \\
512 & 0.2317 & 0.2983 &1.77&0.6925&1.4737&1.35 \\
1024& 0.0725 & 0.1298 &1.68&0.2485&0.9461&1.49 \\
\hline
128 & 3.3427 & 2.4634 &    &5.5060&5.1819&    \\
256 & 0.8539 & 0.6580 &1.97&1.4317&1.8465&1.94\\
512 & 0.2106 & 0.2105 &2.02&0.3775&0.9995&1.92\\
1024& 0.0592 & 0.1036 &1.83&0.1115&0.7909&1.76
\end{tabular}
\end{center}
\end{table}

\subsection{Tests of point force calculation}

In the previous subsection, we have tested different approaches for disk
self-gravity computed at
the cell centers. In this subsection, we describe and compare methods to
calculate the gravitational force 
of the disk on the planet at any point $(r_p,\phi_p)$. We
assume a unit planet mass at that point. 

We investigated the force calculations in three different ways.  
In the first approach, we
performed a direct summation of $\mathcal{G}_r$ and
$\mathcal{G}_\phi$ from the point $(r_p,\phi_p)$ to each of the cell centers
[Eqs. (\ref{Gr1}) and (\ref{Gphi1}) with $(r,\phi) =
(r_p,\phi_p)$]. This is similar to a particle method and is widely
used in calculating the disk force exerted on the planet in the
disk-planet simulations. In the second approach, we
applied bilinear interpolation to the force computed at
the cell centers. In the third approach, we applied a finite difference formula
directly to $\Psi$ at the grid cells closest to the point location.
This approach is very similar to bilinear interpolation of the
force, only it uses a smaller grid stencil, which gives better
accuracy.

To calculate the force at a specific point and
account for the fact that the point can be at an arbitrary location 
within a cell, we take a cell that contains point
$(r_p,\phi_p)=(0.99,0.0)$, which is close to the center of one of
the first Gaussian sphere, and compute the force
for $11\times11$ equidistributed points within the cell. 
Since the third approach is similar to the second
approach and verified to be more accurate, we only compare the first
and third approaches.

\begin{figure}[htbp]
\begin{center}
\includegraphics[width=0.45\textwidth]{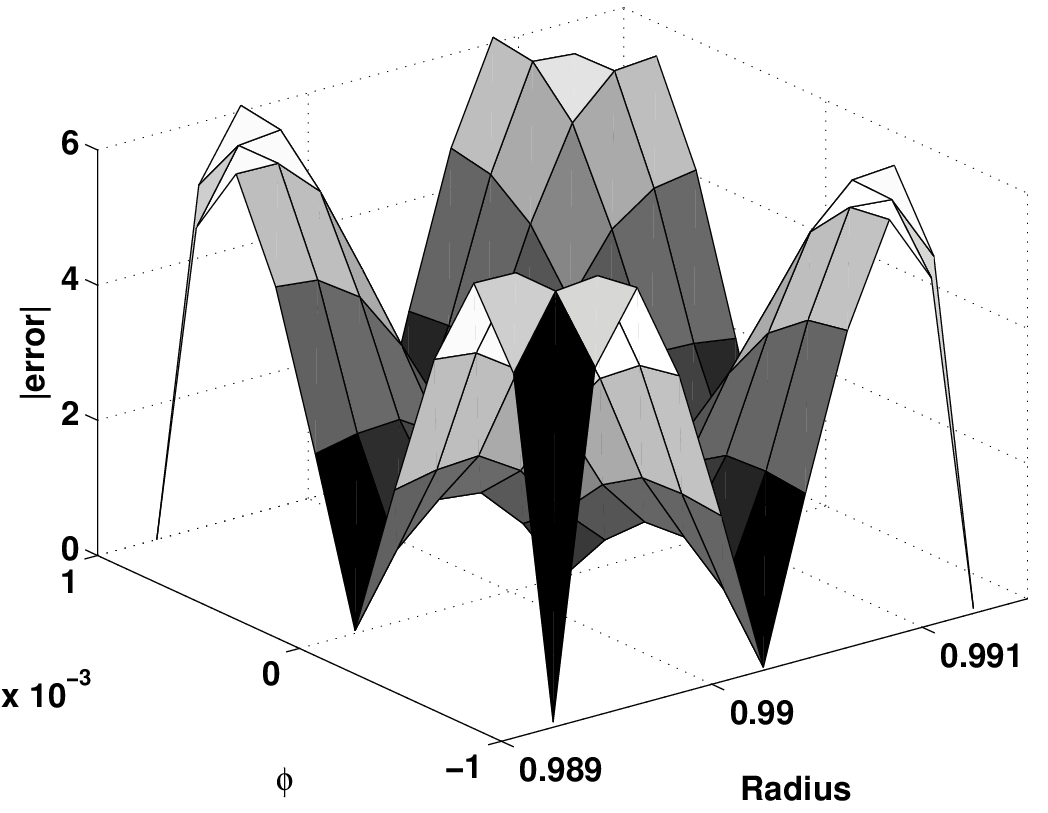}
\includegraphics[width=0.45\textwidth]{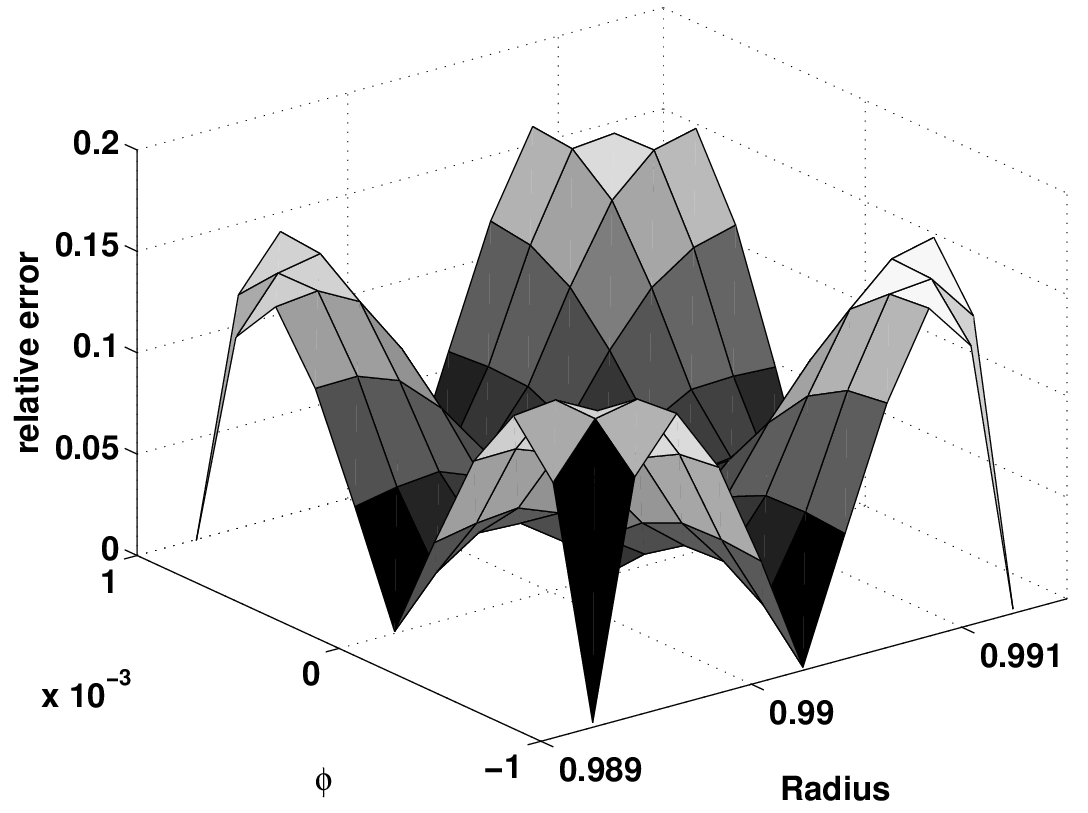}
\includegraphics[width=0.45\textwidth]{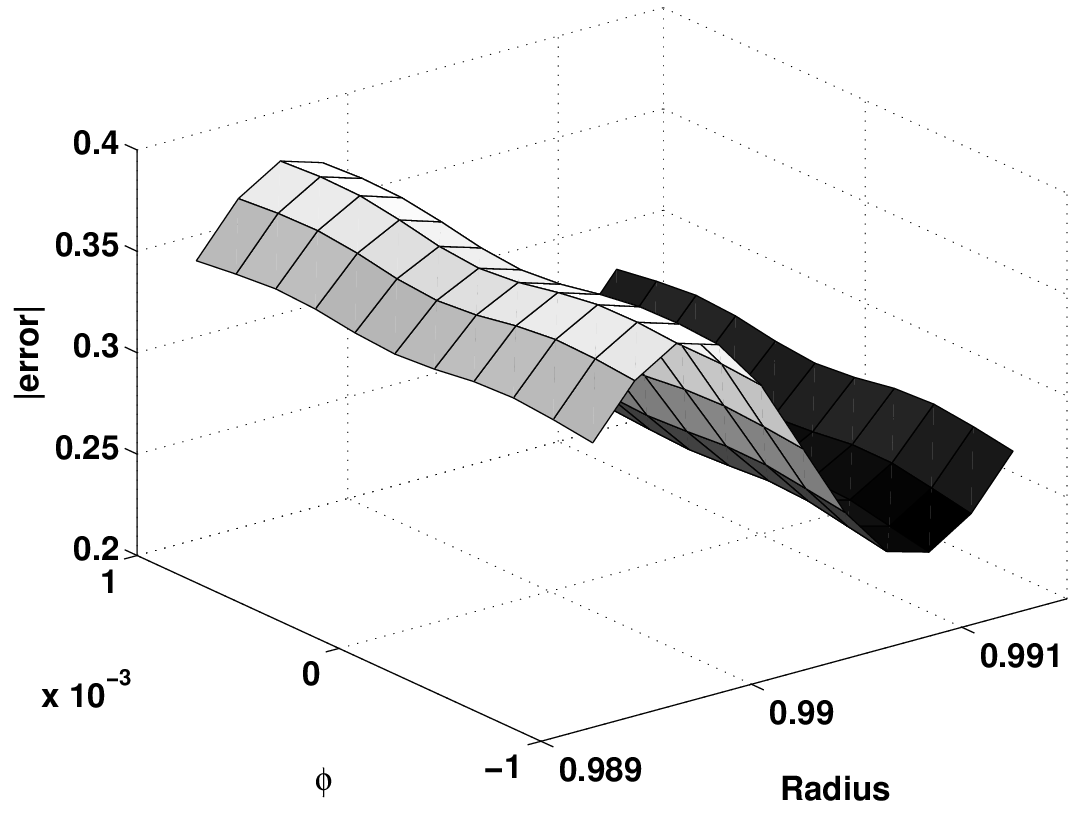}
\includegraphics[width=0.45\textwidth]{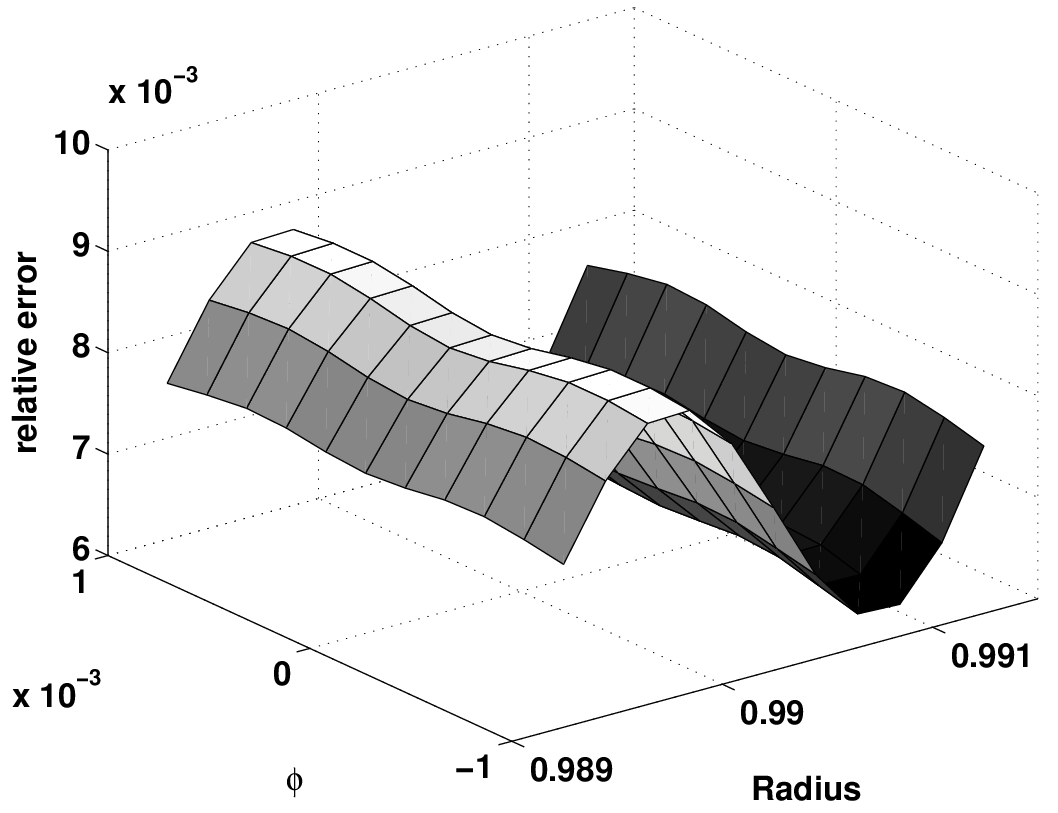}
\includegraphics[width=0.45\textwidth]{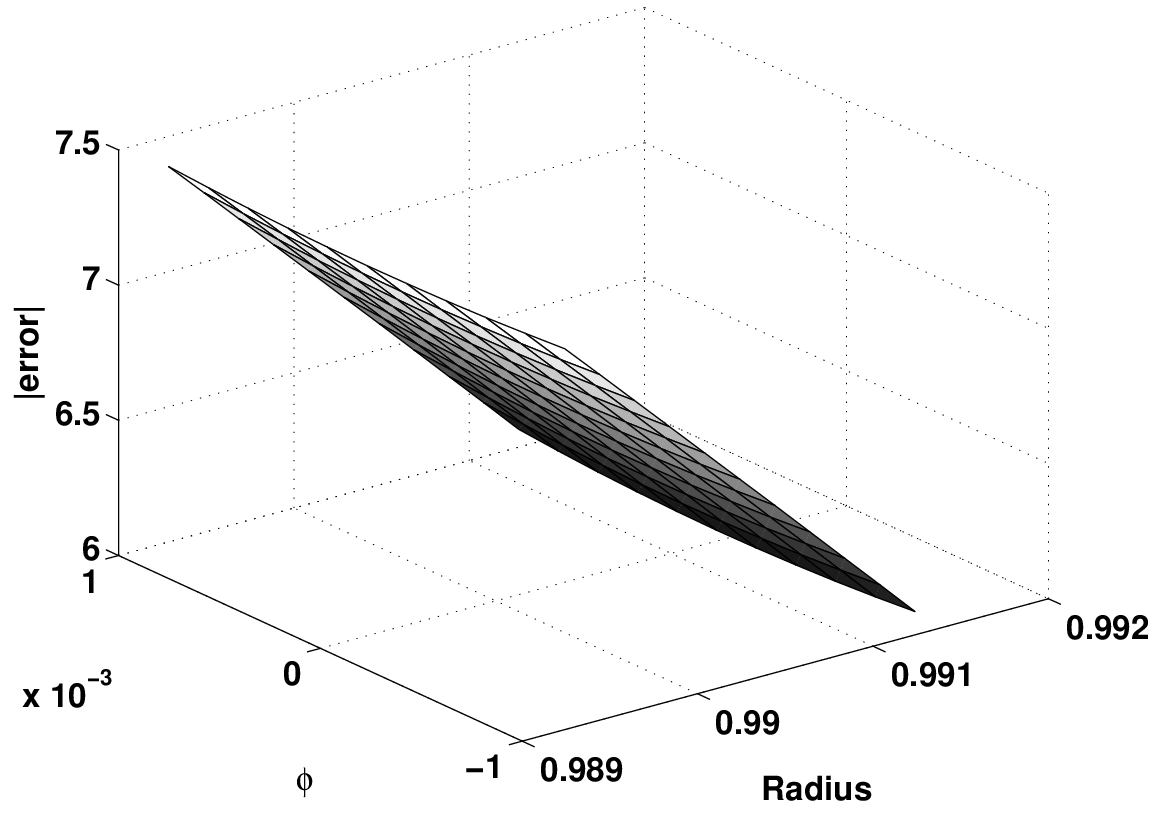}
\includegraphics[width=0.45\textwidth]{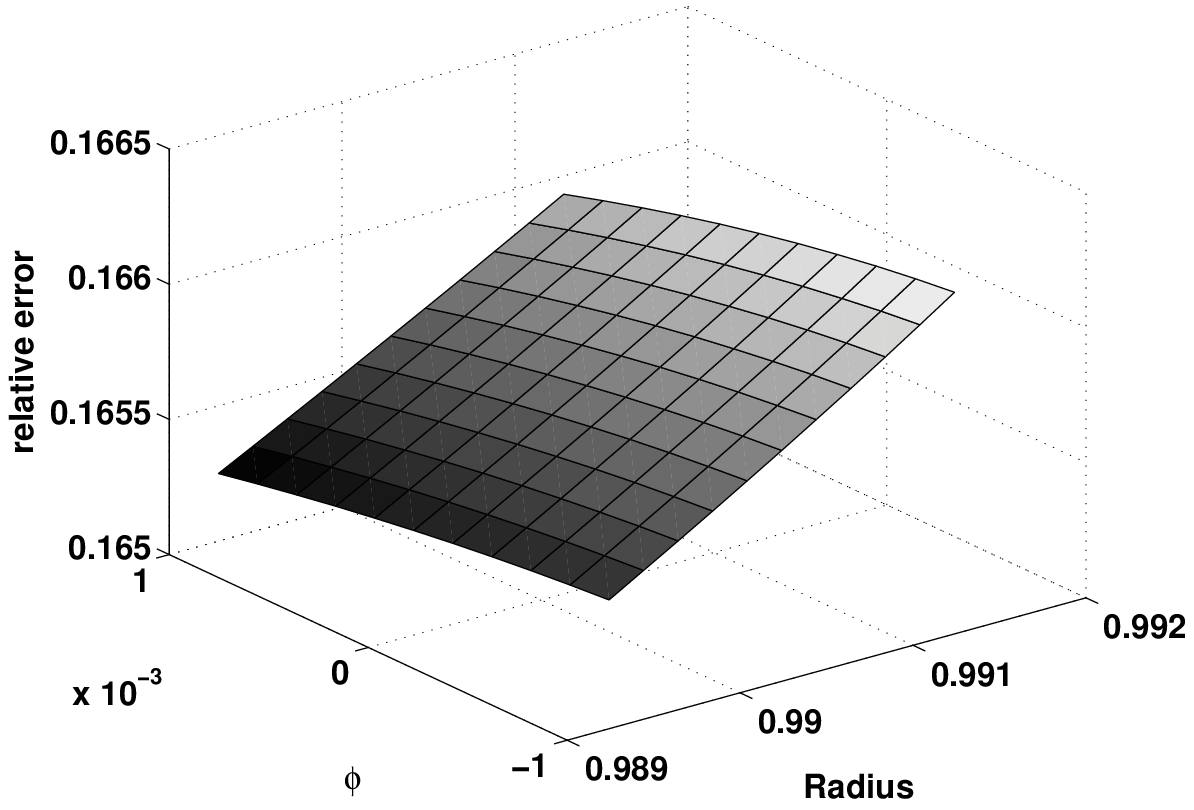}
\caption{\label{pforce1}The accuracy of point force calculation by direct
  summation approach with different softenings: softening (\ref{softn})
  (top two), $\varepsilon = \min(\Delta r,r\Delta\phi)$ (middle two),
  and $\varepsilon = 0.3H(r)$ (bottom two). The left plots show
  the magnitude of the errors and the 
  right plots show the relative errors, both throughout one grid cell.
}
\end{center}
\end{figure}

We use the softening defined by (\ref{softn}) in the direct summation
approach to 
avoid singularities. Fig. \ref{pforce1} shows the
absolute error and relative error throughout the grid cell. We
find that the maximum error and maximum relative error are 6.12 and
16.6\%, respectively. The top two plots of Fig. \ref{pforce1} show
that, although softening (\ref{softn}) works very well for the disk
self-gravity calculated at the cell-center and point force at the
node- and edge-centers, it gives large error for the point force at
other locations. The relative error range is [0.0152\%,16.6\%],
which means the direct summation with this softening is very
sensitive to the location of the point. We have tried two other different
softenings. The middle two plots of Fig. \ref{pforce1} show the
results with softening $\varepsilon = \min(\Delta r,r\Delta\phi)$, which
approximates one grid spacing. Both the maximum error and maximum
relative error are much reduced. Also the range of the relative error
becomes [0.63\%,0.88\%], which means the direction summation with
this softening is relatively insensitive to the location of the
point. The bottom two plots of Fig. \ref{pforce1} show the results
with softening $\varepsilon = 0.3H(r)$, which has been adopted by
\cite{BaMa08} and corresponds to seven
grid spacings in our $800\times3200$ grid layout. Both the maximum
error and relative error become 
large everywhere. The error range is [6.04,7.40], and the relative
error range is [16.5\%,16.6\%].

\begin{figure}[htbp]
\begin{center}
\includegraphics[width=0.45\textwidth]{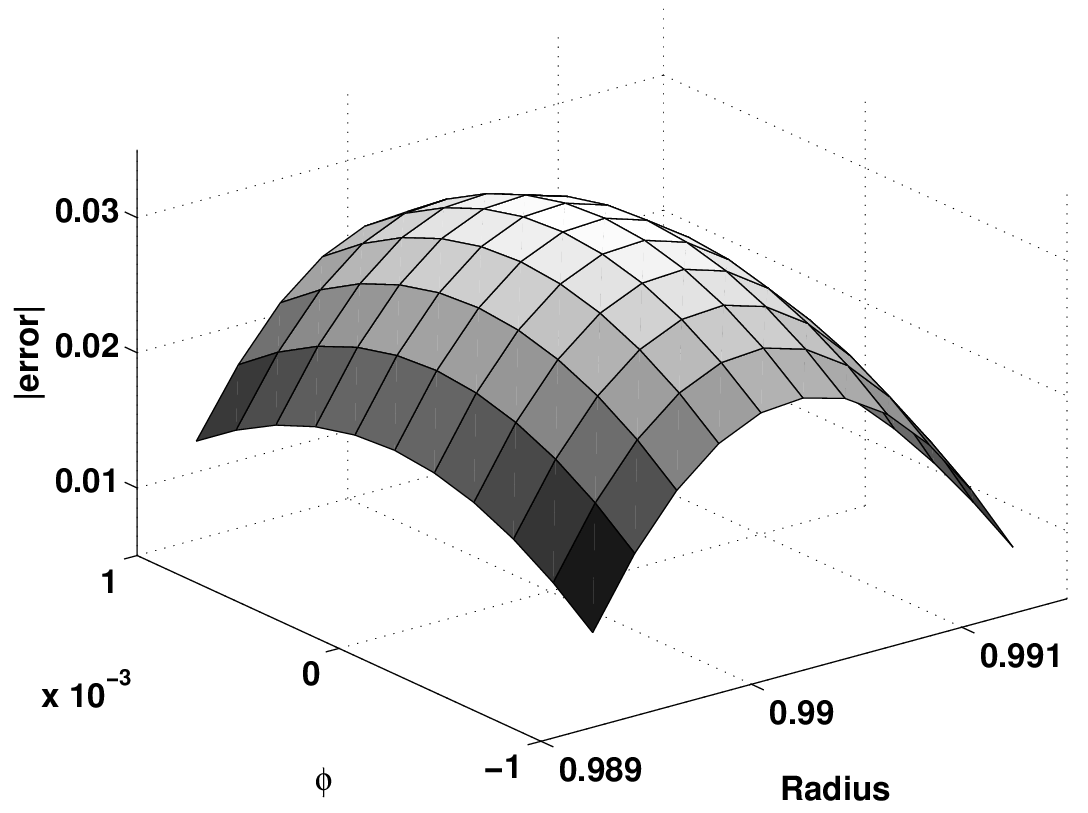}
\includegraphics[width=0.45\textwidth]{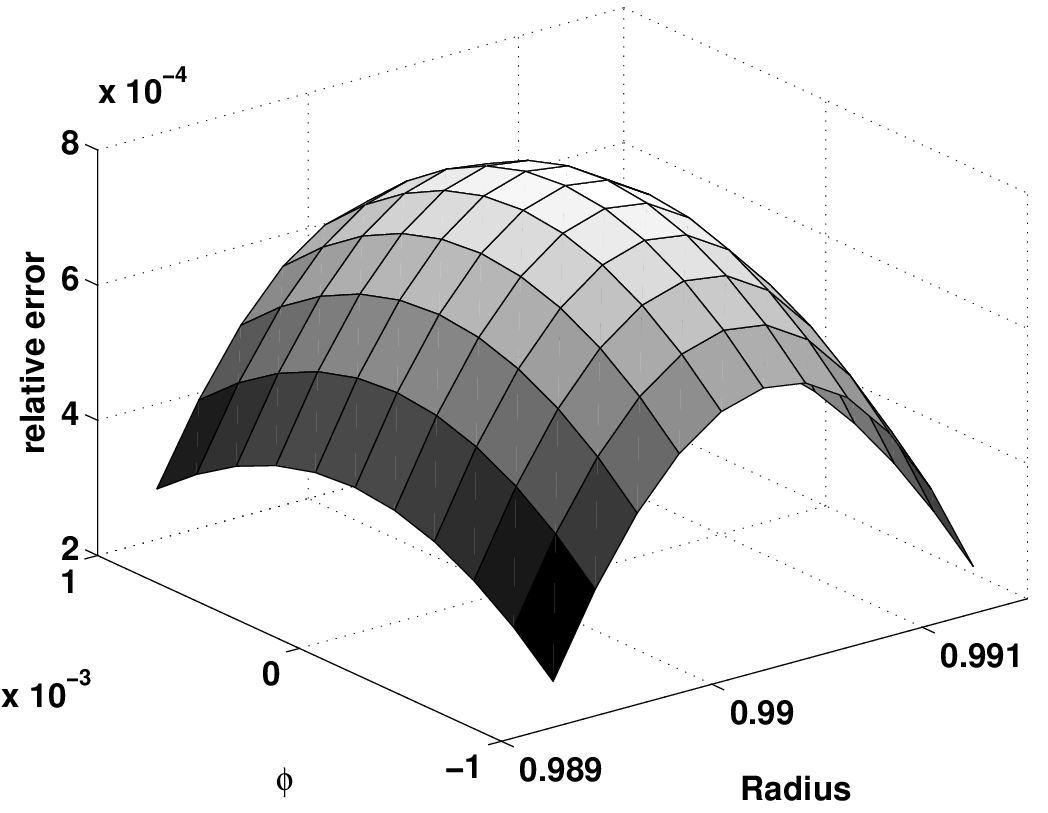}
\caption{\label{pforce2}The accuracy of point force calculation 
  finite-difference of the potential. The left plot show
  the magnitude of the errors and the 
  right plot show the relative errors, both throughout one grid cell.
}
\end{center}
\end{figure}

The third approach using finite difference on 
$\Psi$ achieves much high accuracy, with maximum error and maximum
relative error of 0.0315 and 0.078\%. Fig. \ref{pforce2} shows the
absolute error and relative error throughout the grid cell. The
relative error range is [0.0264\%,0.0778\%], which means that this
approach is
relatively insensitive to the location of the point.

We remark that in the actual disk-planet interaction simulations, 
the accuracy of the force on the planet also depends on how accurately
the disk density around the planet is resolved. Since the disk within
one Roche lobe is usually not well resolved, the third approach in the
point force calculation, though by itself is very accurate, may not 
always give the most ``accurate'' force calculation if the disk density
is poorly determined. In fact, we find that the direct summation approach 
with a relatively large softening, e.g. $\varepsilon=0.1H(r)$, gives results
that are in good agreement with the linear theory results by \cite{Tanaka02}.

\subsection{Tests of the tree-code force calculation}

Here, we test the performance of our tree-code method described in
Section \ref{sec:tc}. First, we test the accuracy with the same
error and convergence measures used in section \ref{sec:test1}. For
simplicity, we use a fixed softening $\varepsilon = \alpha(1)\Delta r
= 0.23\Delta r$ throughout the grid. The approximate model
(\ref{eq_tc7}) is used to calculate 3D correction factor. To
obtain the field in the $z=0$ plane, we set $H=0$ in Eq.
(\ref{eq_tc3}) resulting in an interaction force consistent with the
Green's function method. Also, we use the same density function with
known analytical force field described in Section \ref{sec:test}.
Second, we test the parallel efficiency  by varying the number of
processors for a highly resolved grid (800x3200).

\begin{table}[htbp]
\begin{center}
\caption{\label{tc_fixed_NP} Accuracy and efficiency data for
tree-code force calculation on $N_P$=100 processors.}
\begin{tabular}{cc|ccc|ccc}

Grid &$F_{max}$ &$E_{\max}$ & $RE$($10^{-2}$) & $p$ & total CPU Time
(s)  \\ 
\hline
100 x 400   & 171.68 & 5.617 & 1.5216 &         & 0.1078  \\
200 x 800   & 171.72 & 2.324 & 0.8452 & 1.27    & 0.6844 \\
400 x 1600  & 171.73 & 1.390 & 0.6291 & 0.74    & 3.7167  \\
800 x 3200  & 171.74 & 1.114 & 0.5463 & 0.32    & 16.543  \\
\hline
800 x 3200$^*$  & 171.74 & 0.08912 & 0.0126 &     & 0.02093 
\end{tabular}

{($^*$The last line is for Green function method, where the CPU
  time does not include the pre-computing time, which takes 1.001 second.)}
\end{center}
\end{table}

\begin{table}[htbp]
\begin{center}
\caption{\label{tc_fixed_grid} Parallel efficiency data for
tree-code force calculation on 800x3200 fixed grid.}
\begin{tabular}{c|cc|c}

   & \multicolumn{2}{c}{CPU Time} &  \\
$N_p$ & total time (s) & comp. time (s) & Parallel efficiency \\
\hline
10 & 182.32 & 180.97 & 0.99  \\
20 & 92.78 & 91.83 & 0.98   \\
40 & 46.92 & 46.06 & 0.96   \\
50 & 37.71 & 36.85 & 0.96   \\
80 & 24.07 & 23.05 & 0.94   \\
100 & 20.07 & 18.94 & 0.90  \\
160 & 13.80 & 12.25 & 0.82  \\
200 & 11.28 & 9.49 & 0.80   \\

\end{tabular}
\end{center}
\end{table}

From the results given in Table \ref{tc_fixed_NP} we draw the following
conclusions.  First, this method does not appear to converge to the
exact solution.  This is actually expected since the model force
function (Eq. (\ref{eq_tc7})) with our chosen softening does not
become more accurate as the 
grid is refined.  We observe that the maximum error 
asymptotically approaches to $1.0$, which is much larger than the error using
the Green's function method for highly resolved grids.
Next, we see that the time
complexity scales approximately with $O(N_rN_\phi \log(N_rN_\phi))$, which is
consistent with the theoretical prediction. Finally, we should remark that
the softening has a large impact on the accuracy of the solutions for
the tree-code. If $\varepsilon = 0.3H(r)$ is used, the maximum error
and global relative error become 20.57 and 3.045\% respectively.

Parallel efficiency results can be seen in Table \ref{tc_fixed_grid}.
Although we find our tree-code scales well with the number of
processors, the actual value of the CPU time is more than two orders
of magnitude greater than our Green's function method! We have
experimented with tuning tree code parameters and other memory
optimizations and have found that while it may be possible to gain a
factor of $2-4$ in speedup over the results of Table
\ref{tc_fixed_grid}, our tree-code method always performs {\it much}
slower than our Green's function method. 

\section{Conclusion\label{sec:conc}}

In this paper, we have presented a fast and accurate solver to
calculate the potential and self-gravity forces for the disk systems.
This method is implemented on a polar grid, and FFT can be used in the
azimuthal direction. The pre-calculation of the Green's function and
its FFT play a major role in the algorithm to reduce the computational
cost. We think it could be the main reason why the Green's function
method is much faster than the particle tree-code method.  
We also presented an efficient method in implementing the solver
on parallel computers. We find the computational cost for the
self-gravity solver to be comparable to that of the hydro solver for
large, highly resolved grids run on a distributed memory parallel
architecture.  We also developed a 2D tree-code solver, which uses a
relatively inexpensive model force to accurately account for the
vertical structure of the disk.  

Finally, we notice that if the disk vertical structure varies with
time, our pre-calculation must be done every time step making our
solver inefficient.

We have applied our self-gravity solver to simulations of
disk-planet interaction system. Compared with the simulations without
self-gravity, the total computation time is increased by only 30\% for
a parallel computation with 100 processors. We have also confirmed
that the 2D self-gravity indeed accelerates the planet migration.
These results will be reported elsewhere.

\bigskip
\noindent
{\bf Acknowledgment:}
We would like to thank Dr. C.-K. Chan for helpful discussion during
his stay at Los Alamos.  We also thanks the referee for many useful
comments. This research was performed under the
auspices of the Department of Energy. It was supported by the
Laboratory Directed Research and Development (LDRD) Program at Los
Alamos. It is also available as Los Alamos National Laboratory Report,
\laur{07-5882}.


\end{document}